\documentclass{jfm}
\usepackage[version=4]{mhchem}
\usepackage{amsmath}
\usepackage{graphicx}
\usepackage{epstopdf, epsfig}
\usepackage{amsfonts}
\usepackage{amssymb}
\usepackage{float}
\usepackage{xcolor}
\usepackage{booktabs}
\usepackage{siunitx} 

\usepackage{bm}
\usepackage[acronym]{glossaries}
\usepackage{soul}
\usepackage{tikz}
\usetikzlibrary{shapes,arrows}
\usetikzlibrary{positioning}
\usetikzlibrary{external}
\usepackage{makecell} % add in preamble
\usepackage{enumitem} % needed for customizing lists

\usepackage{dcolumn}% Align table columns on decimal point
\usepackage{bm}% bold math
\usepackage{hyperref}
\usepackage{multirow}
\usepackage{comment}
\usepackage{xcolor}
\usepackage{tikz}
\usetikzlibrary{arrows,decorations.markings,shapes,positioning,arrows.meta}

\tikzset{
  block_m/.style ={rectangle, draw=black, thick, fill=white,
    text width=14em, text centered, minimum height=3em},
  block_s/.style ={rectangle, draw=black, thick, fill=white,
    text width=8em, text centered, minimum height=3em},
  >={Stealth}
}
\tikzset{
  |-|/.style={
    to path={
      -- ++(0,-0.5)      % go vertically down
      -- ++(\x,0)        % go horizontally
      -- ++(0,0.5)       % go vertically up
      -- (\tikztotarget) % connect to target
    }
  }
}
\tikzset{
  -|-/.style args={#1 and #2}{
    to path={
      -- ++(#1,0) -- ++(0,\y) -- ++(#2,0) -- (\tikztotarget)
    }
  }
}

\usepackage{todonotes}

\usepackage{color,soul}
\usepackage[normalem]{ulem}
\usepackage{soul}

\usepackage{nomencl}
\makenomenclature

\shorttitle{Lattice Boltzmann model for non-ideal compressible fluid dynamics}
\shortauthor{S. A. Hosseini, M. Feinberg, I. V. Karlin}

\title{Lattice Boltzmann model for non-ideal compressible fluid dynamics}

\author{S. A. Hosseini\aff{1\corresp{\email{}shosseini@ethz.ch}},
 M. Feinberg\aff{{1\corresp{\email{}mfeinberg@ethz.ch}}},
  I. V. Karlin\aff{{1\corresp{\email{}ikarlin@ethz.ch}}}
}
\affiliation{
\aff{1}Department of Mechanical and Process Engineering, ETH Zurich, 8092 Zurich, Switzerland.}

\begin{document}
\maketitle
\begin{abstract}
We present a new kinetic model and its lattice Boltzmann realization for the simulation of compressible, non-ideal fluid flows. The method employs first-neighbour lattices and introduces a consistent set of correction terms constructed via quasi-equilibrium attractors, ensuring positive-definite and Galilean-invariant Navier–Stokes dissipation rates. This construction circumvents the need for extended stencils or ad hoc regularization, while maintaining numerical stability and thermodynamic consistency across a broad range of flow regimes. The resulting model accurately reproduces both the Euler- and Navier--Stokes hydrodynamic limits. As a stringent validation, we demonstrate, for the first time within a lattice Boltzmann framework, quantitatively accurate simulations of shock-–drop interactions at Mach numbers up to 1.47. The proposed approach thus extends the applicability of lattice Boltzmann methods to high-speed, non-ideal compressible flows with a minimal kinetic stencil.
\end{abstract}


%\printnomenclature

\section{\label{sec:introduction}Introduction}
\begin{figure}
\centering
\includegraphics[keepaspectratio=true,width=0.4\textwidth]{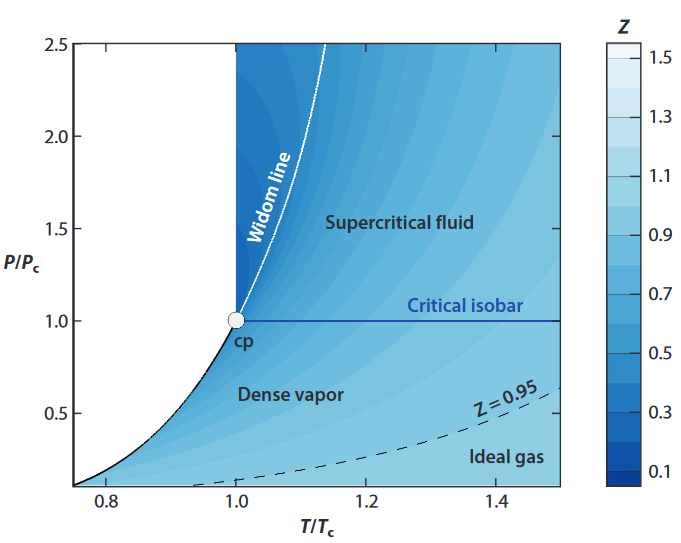}
  \caption{Pressure-temperature diagram for $\ce{CO2}$. The colour scale indicates the compressibility factor $Z=P/\rho RT$. Figure reproduced from~\cite{guardone2024nonideal}. Here $P_c$ and $T_c$ are the critical pressure and temperature.}
 \label{fig:CO2_diagram}
\end{figure}

Non-ideal compressible fluid dynamics is a novel and rapidly developing branch of fluid mechanics, mainly due to the emergence of methods and technologies operating in the near-, trans- and super-critical regimes. It is, in part, concerned with the gas-dynamics of single-phase fluids in non-ideal thermodynamic states, i.e., states where the compressibility factor differs from unity~\citep{guardone2024nonideal}. These states are illustrated in a pressure--temperature diagram for ${\rm CO}_2$ in Figure~\ref{fig:CO2_diagram}.
Non-ideal compressible fluid dynamics encompasses the dynamics of supercritical fluids, dense vapours, and liquids. 
The latter two are separated by the Widom line, an extension beyond the critical point that delineates \emph{vapour-like} and \emph{liquid-like} behaviour, identified by maxima in the constant-pressure specific heat \citep{simeoni2010widom}.
In addition, it includes fluids of higher molecular complexity with a negative fundamental derivative of gas dynamics~\citep{colonna2009computation}, also known as Bethe-Zel’dovich-Thompson (BZT) fluids~\citep{zel1946possibility,thompson1973negative} as well as simple fluids in the proximity of the critical point. The rapidly growing interest within the energy industry in fluids operating in such thermodynamic states, together with their marked departure from ideal-gas behaviour, demonstrates the need for systematic studies of such flows. However, as evident in the literature, while considerable effort has been devoted to developing experimental set-ups for non-ideal fluid dynamics in recent years~\citep{lettieri2015investigation,zocca2019experimental,gallarini2021direct,head2022mach}, experimental data remains scarce and {is} complicated to acquire~\citep{guardone2024nonideal}. Consequently, the development of consistent and efficient numerical tools for the simulation of such flows is essential for advancing both the fundamental understanding of non-ideal compressible fluid dynamics and technologies such as organic Rankine cycles and supercritical ${\rm CO}_2$ turbines~\citep{guardone2024nonideal,gunawan2023design,chen2023design}.
More specifically, direct numerical simulations are necessary both to understand the complex physics of non-ideal compressible fluid dynamics and to generate engineering databases. Examples include skin friction and dissipation coefficients~\citep{cramer1996transonic,pini2018entropy}, heat transfer in wall-bounded turbulent supercritical flows~\citep{peeters2016turbulence,kawai2019heated}, and the critical mass flow rate and pressure in nozzles and turbines involving flashing flows.
While direct numerical simulations may at times be cost-prohibitive for full-scale simulations, they are a necessary tool for the development of subgrid-scale closure models required for the more cost effective large eddy and Reynolds-averaged Navier--Stokes simulations~\citep{garnier2009large}.
Recent studies on homogeneous isotropic turbulence fluid structures in dense vapours have shown that while large-scale flow structures are mostly affected by molecular complexity, smaller-scale flow structures are impacted by local variations in sound speed~\citep{sciacovelli2017small,duan2021dense}. In addition, the dense vapour has been shown to display modified shocklet structures, with considerably reduced jumps in pressure, density, and entropy in compression shocklets, and the emergence of expansion shocklets~\citep{giauque2017direct,giauque2020direct}. Similarly, substantial differences from ideal gas dynamics have been reported~\citep{vadrot2021direct} for wall-bounded turbulence, particularly at smaller scales, and where strong normal gradients in density and viscosity -- and consequently the local speed of sound and Mach number -- may strongly impact turbulent structures~\citep{pecnik2017scaling,peeters2016turbulence}. Reliable direct numerical simulation studies of non-ideal compressible fluid dynamics can fill existing gaps in the understanding of physics of such flows. A reliable numerical scheme presupposes a physical model valid across all regimes of interest here, i.e. super-, trans- and near-critical flows involving pronounced non-ideal and compressibility effects. 
The development and validation of efficient numerical schemes for Navier-Stokes-Korteweg (NSK) dynamics, regardless of method, can have a significant impact on the broader literature and understanding of non-ideal compressible flows.\\
{The lattice Boltzmann method (LBM) is an excellent candidate for the flow regimes of interest here.
Already well established as an efficient method for resolved simulation of flows in the incompressible limit, the distinct numerical features of LBM offer advantages for compressible, non-ideal fluid dynamics.
Specifically, LBM benefits from the decoupling of non-linearity and non-locality noted by S. Succi, in LBM ``nonlinearity is local, non-locality is linear''~\citep{succi2001lattice,succi2018lattice}.
LBM solvers are conservative and low-dissipative schemes with spectral properties -- both dissipative and dispersive -- that compare favorably to conventional solvers of the same order~\citep{hosseini2020development,hosseini2019extensive,hosseini2022lattice,martinez1994comparison,wissocq2019extended,peng2010comparison}, especially for normal propagation modes, i.e. acoustic modes~\citep{bres2009properties,viggen2011viscously,viggen2014acoustic}. Along with incompressible ideal fluid dynamics, two-phase flow simulation has witnessed major success and growth in popularity in LBM, starting with formulations such as the colour-gradient~\citep{gunstensen1991lattice}, pseudo-potential~\citep{shan1993lattice} and free energy~\citep{swift1996lattice} in the early 1990s. The latter two are of special interest in the context of non-ideal fluid dynamics as they effectively solve a form of the NSK. The vast majority of the literature that focuses on or uses this class of models has been tailored to become essentially highly efficient interface tracking models for multi-phase flows~\citep{luo2021unified,chikatamarla2015entropic,hosseini2023lattice,chen2014critical,li2016lattice}. Non-ideal compressible thermodynamics have long been neglected in LBM applications. The free energy model, for instance, which is the only one out of the previously-listed approaches that results from minimization of a free energy functional under a constraint on the total mass, can readily be shown to be related to mean-field kinetic models such as the Boltzmann--Enskog--Vlasov~\citep{hosseini2022towards,hosseini2023lattice} equations and recover, in the hydrodynamic limit and under specific scaling, the NSK equations~\citep{hosseini2022towards} and mean-field van der Waals fluid thermodynamics. This means that it cannot only model two-phase flow dynamics and phase separation but also properly recover all mean-field, near-, trans-, and super-critical behaviours associated with non-ideal fluids, as demonstrated, for instance, through studies of properties such as the Tollman length~\citep{hosseini2022towards,hosseini2023lattice,reyhanian2020thermokinetic,reyhanian2021kinetic,lulli2022mesoscale,lulli2022metastable,hazi2008modeling}.\\
The use and extension of this class of thermodynamically consistent models to compressible non-ideal fluid dynamics is a largely under-explored area that can considerably impact research on non-ideal compressible fluid dynamics. Early attempts at developing an LBM for compressible non-ideal flows were documented in \citep{he2002thermodynamic}. Since then, thermal multi-phase models based on the pseudo-potential approach have witnessed steady growth. However, the vast majority of the models and studies in the literature have been tailored to boiling applications, see for instance \citep{li2015lattice,fei2020mesoscopic,saito2021lattice,fang2017lattice,huang2021mesoscopic}. To the author's knowledge, the only documented attempts at modeling compressible non-ideal flows beyond evaporation are \citep{vienne2024hybrid}, where a hybrid lattice Boltzmann/finite-volumes scheme -- with a finite-volume discretization of the energy equation -- was proposed, and \citep{reyhanian2020thermokinetic,reyhanian2021kinetic}, where the authors demonstrated a numerical model based on the Particles-on-Demand realization of the lattice Boltzmann method~\cite{dorschner2018particles}. We propose to address this gap in the literature with {a novel kinetic model and its} lattice Boltzmann {realization} for non-ideal fluids in the compressible regime. To retain the main advantages of the lattice Boltzmann method, the proposed model relies on classical first-neighbour lattices, taking advantage of a second distribution function for the energy balance equation~\citep{rykov1975model,saadat2021extended,prasianakis2007lattice,prasianakis2008lattice}. {The kinetic model relies on a novel set of BGK-like collision operators with local- and shifted-equilibrium attractors ensuring recovery of target hydrodynamics. For instance independent control over the bulk viscosity is guaranteed through the shifted equilibrium;} The latter is a critical point as the bulk viscosity dictated by the BGK structure, for equilibria with pressures other than the ideal gas pressure, can take on negative values in the hydrodynamic limit; see \citep{hosseini2022towards}. Together with a consistent second-order discretization and treatment of source terms, the model will be shown to correctly recover the target hydrodynamic limit.\\
The paper is organized as follows: Section~\ref{sec:2} introduces the target hydrodynamic limit. Section~\ref{sec:LBM} presents the kinetic model and its lattice Boltzmann realisation. Section~\ref{sec:nums} provides validation across a range of increasingly complex configurations and studies of non-ideal fluid-specific dynamics through cases such as shock tubes and shock--liquid-column interaction. The article closes with final remarks in Section~\ref{sec:level4}.
\section{Balance equations for a compressible non-ideal fluid\label{sec:2}}
We begin with a brief overview of a one-component, compressible non-ideal fluid system.
Material presented in this section is standard and serves to define the target hydrodynamics for a kinetic model to be introduced in section \ref{sec:LBM}.
We introduce the macroscopic fields of fluid density $\rho(\bm{x},t)$, momentum $\rho\bm{u}(\bm{x},t)$, and \emph{bulk energy} $\rho E(\bm{x},t)$. The latter is the sum of the flow kinetic-energy density and the internal-energy density $\rho e$,
{\begin{equation}\label{eq:rhoE}
    \rho E = \rho{e}+\frac{1}{2}\rho u^2.
\end{equation}
}
The specific internal-energy $e(v,T)$ per unit mass is a function of absolute temperature $T$ and specific volume $v=1/\rho$, and is defined by a familiar thermodynamic relation for its differential,
\begin{equation}\label{eq:de}
{de=c_v\,dT+\left[T\left(\dfrac{\partial P}{\partial T}\right)_v-P\right]{dv},}
\end{equation}
{where $P(v,T)$ is the thermodynamic pressure and $c_v$ is the specific heat at constant volume,
\begin{equation}
    c_v=\left(\dfrac{\partial e}{\partial T}\right)_v.
\end{equation}
In the following, it will be convenient to consider the thermodynamic equation of state as a function of the density $\rho$ rather than of the specific volume $v$,
\begin{equation}
P(\rho,T)=\left.P(v,T)\right|_{v={1}/{\rho}},
\end{equation}
so that the differential of the internal energy \eqref{eq:de} becomes
\begin{equation}\label{eq:de_rho}
{de=c_vdT-\left[T\left(\dfrac{\partial P}{\partial T}\right)_\rho-P\right]\frac{d\rho}{\rho^2}.}
\end{equation}}

The mass, momentum and bulk energy balance equations are,
    \begin{align}
       & \partial_t \rho + \bm{\nabla}\cdot \rho\bm{u} = 0,
    	\label{eq:balance_density}
    	\\
    	&\partial_t (\rho\bm{u}) + \bm{\nabla}\cdot \rho\bm{u}\otimes\bm{u} + \bm{\nabla}P + \bm{\nabla}\cdot \bm{T}_{\rm NS} + \bm{\nabla}\cdot \bm{T}_{\rm K}= 0,
    	\label{eq:balance_momentum}\\
        &{\partial_t (\rho E) + \bm{\nabla}\cdot \left(\rho E\bm{u}+{P}\bm{u}\right) + \bm{u}\cdot\left(\bm{\nabla}\cdot \bm{T}_{\rm K}\right) + \bm{\nabla}\cdot(\bm{u}\cdot\bm{T}_{\rm NS}) + \bm{\nabla}\cdot{\bm{q}_{\rm F}}= 0.}
    	\label{eq:balance_energy}
    \end{align}
\nomenclature{$\bm{u}$}{Fluid velocity vector}
The viscous stress tensor is defined as
\begin{equation}\label{eq:TNS}
    \bm{T}_{\rm NS} = {-}\mu\left(\bm{\nabla}\bm{u} + \bm{\nabla}\bm{u}^\dagger - \frac{2}{D}(\bm{\nabla}\cdot\bm{u})\bm{I}\right) - \eta (\bm{\nabla}\cdot\bm{u})\bm{I},
\end{equation}
{where $\mu$ and $\eta$ are the dynamic shear and bulk viscosity coefficients, respectively.}
Furthermore, the Korteweg surface tension tensor is defined as
\begin{equation}
    \bm{T}_{\rm K} = \kappa \bm{\nabla}\rho\otimes\bm{\nabla}\rho - \kappa\left(\rho (\bm{\nabla}\cdot\bm{\nabla})\rho+\frac{1}{2}{\lvert \bm{\nabla}\rho\rvert}^2\right)\bm{I},
\end{equation}
{where $\kappa$ is the capillarity coefficient.}
{Since only the divergence of Korteweg's tensor contributes to the balance equations, we define the Korteweg's force as
\begin{equation}\label{eq:FK}
    \bm{F} = -\bm{\nabla}\cdot \bm{T}_{\rm K}=\kappa\rho\bm{\nabla} (\bm{\nabla}\cdot\bm{\nabla})\rho.
\end{equation}
}
Finally, the Fourier heat flux is defined as
\begin{equation}\label{eq:qF}
    {\bm{q}_{\rm F}} = -k\bm{\nabla}T,
\end{equation}
where $k$ is the thermal conductivity coefficient. All transport coefficients are considered constants below.

While the bulk energy balance equation \eqref{eq:balance_energy} is our primary target equation in the following, we mention several other related forms that are implied by the system \eqref{eq:balance_density}, \eqref{eq:balance_momentum}, and \eqref{eq:balance_energy}. 
Denoting by $\mathcal{K}=(1/2) \rho u^2$ the flow kinetic energy, and computing its time derivative using the continuity equation \eqref{eq:balance_density} and the momentum balance \eqref{eq:balance_momentum}, we obtain
\begin{equation}
    \label{eq:Kin_energy}
    \partial_t \mathcal{K}+\bm{\nabla}\cdot\left( \mathcal{K}\bm{u}\right)=-\bm{u}\cdot\bm{\nabla}P
    - (\bm{u}\bm{\nabla}):\bm{T}_{\rm K} - (\bm{u}\bm{\nabla}):\bm{T}_{\rm NS}.
\end{equation}
Using the decomposition \eqref{eq:rhoE}, the internal-energy balance is obtained by subtracting the kinetic-energy balance \eqref{eq:Kin_energy} from the bulk-energy balance \eqref{eq:balance_energy}:
\begin{equation}\label{eq:e}
    \rho  (\partial_t e+\bm{u}\cdot\bm{\nabla} e)= -P(\bm{\nabla}\cdot\bm{u}) -\bm{T}_{\rm NS}:\bm{\nabla} \bm{u}
-\nabla\cdot \bm{q}_{\rm F}.
\end{equation}
Furthermore, using the differential of the internal energy \eqref{eq:de_rho} and the continuity equation \eqref{eq:balance_density}, one obtains the temperature equation,
\begin{equation}\label{eq:temperature}
    \rho c_v (\partial_t T+\bm{u}\cdot\bm{\nabla} T)=-T \left(\frac{\partial P}{\partial T}\right)_\rho (\bm{\nabla}\cdot\bm{u})
-\bm{T}_{\rm NS}:\bm{\nabla} \bm{u} -\nabla\cdot \bm{q}_{\rm F},
\end{equation}
and similarly, the pressure equation,
\begin{equation}\label{eq:balance_pressure}
    \partial_t P+\bm{u}\cdot\bm{\nabla} P=-\rho c_s^2(\bm{\nabla}\cdot\bm{u})-\frac{1}{\rho c_v}\left(\dfrac{\partial P}{\partial T}\right)_\rho(\bm{T}_{\rm NS}:\bm{\nabla} \bm{u}
+\nabla\cdot \bm{q}_{\rm F}),
\end{equation}
where $c_s$ is the  speed of sound,
\begin{equation}
    \label{eq:speed_of_sound}
    c_s^2={\left(\dfrac{\partial P}{\partial\rho}\right)_T+\frac{T}{\rho^2 c_v}\left(\dfrac{\partial P}{\partial T}\right)_\rho^2}.
\end{equation}
A further useful form is the balance for the \emph{total energy} ${\mathcal{E}}$, which takes into account the energy of the liquid-vapour interface.
\begin{equation}\label{eq:E_full}
     {\mathcal{E}}=\rho E+\frac{1}{2} \kappa|\bm{\nabla}\rho|^2.
\end{equation}
Introducing the interface energy, $\mathcal{E}_{\kappa}=({1}/{2}) \kappa|\bm{\nabla}\rho|^2$, we first compute its time derivative using the continuity equation \eqref{eq:balance_density} to get,
\begin{equation}
    \partial_t \mathcal{E}_{\kappa} + \bm{\nabla}\cdot \left(\mathcal{E}_{\kappa}\bm{u}\right) + \bm{T}_{\rm K}:\bm{\nabla}\bm{u} + \bm{\nabla}\cdot\left(\kappa\rho(\bm{\nabla}\cdot\bm{u})\bm{\nabla}\rho\right)= 0.
    	\label{eq:balance_surface}
\end{equation}
Adding the bulk-energy balance \eqref{eq:balance_energy} and the interface-energy balance \eqref{eq:balance_surface}, we obtain the balance equation for the \emph{total energy} \eqref{eq:E_full} in the standard form,
\begin{equation}\label{eq:balance_total_e}
      \partial_t{\mathcal{E}}+\bm{\nabla}\cdot\left({\mathcal{E}}{\bm{u}}+P{\bm{u}}+{\bm{T}_{\rm NS}}\cdot{\bm{u}}+
          \bm{T}_{\rm K}\cdot{\bm{u}}+\kappa\rho(\bm{\nabla}\cdot{\bm{u}})\bm{\nabla}\rho+\bm{q}_{\rm F}\right)=0.
\end{equation}
In summary, the various energy-balance forms listed above are implied by the set of balance equations for mass \eqref{eq:balance_density}, momentum \eqref{eq:balance_momentum} and bulk energy \eqref{eq:balance_energy}, which constitute the target equations for the lattice Boltzmann realization. While the total energy balance form \eqref{eq:balance_total_e} seems to be preferred for conventional finite-volume methods, the bulk energy version of the energy balance is more convenient in the lattice Boltzmann setting due to the locality of the corresponding bulk energy field.

Furthermore, we mention a special case in which the specific heat at constant volume is a function of the absolute temperature only, $c_v=c_v(T)$. This implies that the thermodynamic equation of state is a linear function of temperature,
\begin{equation}\label{eq:eos_general}
    P(v,T)=Tb(v)+a(v),
\end{equation}
where $a$ and $b$ are functions of specific volume (or density) only. Indeed, since \eqref{eq:de} is a complete differential, the equality of mixed derivatives implies,
\begin{equation}\label{eq:cv_zero}
    \dfrac{\partial c_v}{\partial v}=\dfrac{\partial}{\partial T}\left[T\left(\dfrac{\partial P}{\partial T}\right)-P\right]=T\dfrac{\partial^2P}{\partial T^2}=0.
\end{equation}
Commonly used examples are the van der Waals and the Carnahan--Starling equations of state. 
From a more microscopic viewpoint, corresponding balance equations are derived from the Enskog--Vlasov kinetic equation.
To link the aforementioned energy equations to those appearing in the kinetic theory, we decompose the 
pressure into  hard-sphere and mean field contributions.
We define the excluded volume, or Enskog, contributions as, $P_{\rm hs}=TB(\rho)=Tb(1/\rho)$, and the mean-field, or Vlasov, contributions as $P_{\rm mf}=A(\rho)=a(1/\rho)$.
This permits the above equation of state \eqref{eq:eos_general} to be written,
\begin{equation}
    \label{eq:partition_PEV}
    P=P_{\rm hs}+P_{\rm mf}.
\end{equation}
The specific internal energy is partitioned accordingly,
\begin{align}
    & de=de_{\rm hs}+de_{\rm mf},\\ 
    & de_{\rm hs}=c_v(T)dT,\label{eq:dehs}\\
    & de_{\rm mf}=P_{\rm mf}\frac{d\rho}{\rho^2}.\label{eq:demf}
\end{align}
Here $e_{\rm hs}$ is the specific thermal energy, and $e_{\rm mf}$ is the specific molecular potential energy (in the mean-field approximation). For the Enskog model of hard spheres, the specific heat is that of the ideal monatomic gas, $c_v=(3/2)R$. The balance equation for the thermal energy $\rho e_{\rm hs}$ is obtained by excluding the molecular potential energy from the balance of the internal energy \eqref{eq:e}.
\begin{equation}
    \label{eq:balance_ehs}
    \partial_t (\rho {e}_{\rm hs}) + \bm{\nabla}\cdot \left(\rho e_{\rm hs}\bm{u}\right) + P_{\rm hs}(\bm{\nabla}\cdot\bm{u})+ \bm{T}_{\rm NS}:\bm{\nabla}\bm{u} + \bm{\nabla}\cdot\bm{q}_{\rm F}= 0.
\end{equation}
Introducing the total kinetic energy of hard spheres,
\begin{equation}
    \label{eq:Ehs}
\rho E_{\rm hs}=\rho e_{\rm hs}+\frac{1}{2}\rho u^2,
\end{equation}
we obtain its balance upon adding the flow kinetic energy equation \eqref{eq:Kin_energy} to the thermal energy balance \eqref{eq:balance_ehs}:
\begin{equation}
\partial_t (\rho {E}_{\rm hs}) + \bm{\nabla}\cdot \left(\rho E_{\rm hs}\bm{u}+P_{\rm hs}\bm{u}\right) + \bm{u}\cdot\left(\bm{\nabla}P_{\rm mf}+\bm{\nabla}\cdot \bm{T}_{\rm K}\right) + \bm{\nabla}\cdot(\bm{u}\cdot\bm{T}_{\rm NS}) + \bm{\nabla}\cdot\bm{q}_{\rm F}= 0.
    	\label{eq:balance_energy_enskog}
        \end{equation}
The balance equation \eqref{eq:balance_energy_enskog} can be derived directly from the Enskog--Vlasov kinetic equation under appropriate scaling in the hydrodynamic limit. Conversely, starting with the energy balance \eqref{eq:balance_energy_enskog} and adding the balance for the potential energy \eqref{eq:demf} (since $e_{\rm mf}$ depends only on the density, the latter follows from the continuity equation),
\begin{equation}
    \label{eq:balance_emf}
 \partial_t (\rho {e}_{\rm mf}) + \bm{\nabla}\cdot \left(\rho e_{\rm mf}\bm{u}+P_{\rm mf}\bm{u}\right) - \bm{u}\cdot\bm{\nabla}P_{\rm mf}
 = 0,  
\end{equation}
we recover the bulk energy balance \eqref{eq:balance_energy} (and, consequently, after adding the balance of the interface energy \eqref{eq:balance_surface}, the total energy balance \eqref{eq:balance_total_e}), for the special case of the equation of state \eqref{eq:eos_general}. These considerations show that a more general phenomenological bulk energy balance \eqref{eq:balance_energy} is consistent with the special case derived directly from kinetic theory and thus gives us a further reason to use \eqref{eq:balance_energy} as the target energy balance equation in the lattice Boltzmann context.
Consequently, and without loss of generality, we use the van der Waals  equation of state in the numerical examples below,
\begin{equation}\label{eq:Pvdw}
    P(\rho,T) = \frac{\rho R T }{1 - b\rho} - a\rho^2.
\end{equation}
\nomenclature{$a$}{Van der Waals long range attractive interaction coefficient}
\nomenclature{$b$}{Van der Waals excluded volume coefficient}
The excluded volume parameter $b$ and the long-range molecular attraction parameter $a$ are defined in terms of the critical-state thermodynamic data: the critical density $\rho_c$, critical temperature $T_c$ and critical pressure $P_c$, as follows:
        $a ={27 R^2 T_c^2}/{64 P_c}$,
    	$b = {R T_c}/{8 P_c}$.
The differentials of the specific internal energy \eqref{eq:de_rho} and of the specific entropy $s$ for the van der Waals fluid are, respectively,
\begin{align}\label{eq:de_vdW}
    &{de={c_v\,dT}-a\,d\rho,}\\
\label{eq:ds_vdW}
    &{ds=\frac{c_v}{T}\,dT-\frac{R}{\rho(1-b\rho)}\,d\rho.}
\end{align}
In the next section, we shall introduce a lattice Boltzmann model that recovers the above system of balance equations \eqref{eq:balance_density}, \eqref{eq:balance_momentum}, and \eqref{eq:balance_energy} in the hydrodynamic limit.
Before doing so, we highlight a motivation for adopting the lattice Boltzmann formulation for modelling compressible non-ideal fluids. From the pressure equation \eqref{eq:balance_pressure}, one observes that, for a non-ideal fluid, the adiabatic speed of sound squared \eqref{eq:speed_of_sound} becomes negative for a van der Waals-type equation of state in the thermodynamically unstable spinodal region of the density-temperature diagram. Consequently, in the inviscid limit, the evolution equation changes type from hyperbolic to elliptic, and special treatment invoking Maxwell's equal-area rule has to be applied in conventional CFD methods. In contrast, the lattice Boltzmann method, by being based in kinetic theory, is able to circumvent this issue as it inherits propagation along fixed characteristics, namely the discrete velocities. Thus, the lattice Boltzmann model introduced below should not be viewed as yet another interface-capturing numerical scheme for multiphase flows but rather as a reduced kinetic theory {targeting the thermodynamically consistent compressible Navier--Stokes--Korteweg hydrodynamic limit}.
\section{Lattice Boltzmann model for non-ideal compressible flows\label{sec:LBM}}
\nomenclature{$\delta t$}{Time-step size}
\nomenclature{$\delta x$}{Grid size}
\nomenclature{$\bm{v}_i$}{Lattice discrete velocities in dimensional units}
\nomenclature{$\bm{c}_i$}{$\bm{v}_i/(\delta x/\delta t)$}

\subsection{Kinetic model}\label{sec:kineticmodel}
{
In this section, we introduce a kinetic model tailored to recover the target hydrodynamic equations of a compressible non-ideal fluid, Eqs. \eqref{eq:balance_density}, \eqref{eq:balance_momentum} and \eqref{eq:balance_energy}, in the hydrodynamic limit. 
To this end, we follow the so-called double distribution function approach and consider two velocity distribution functions,
$f(\bm{v}, \bm{x}, t)$ and $g(\bm{v},\bm{x}, t)$, where $\bm{v}$ is the velocity of a particle. 
The idea of a double distribution function kinetic model was first proposed by \cite{rykov1975model} for polyatomic molecules, where the second distribution function represents the rotational-vibrational contribution to the internal energy. 
This approach was later adopted in the lattice Boltzmann method \citep{he_1998_novel,guo2007thermal,li2007coupled,karlin2013consistent}, where  the second distribution function has been used to represent different forms of energy. 
Here we propose a model where the bulk energy $\rho E$ \eqref{eq:rhoE} is represented by the second distribution function, although other choices are  possible. This specific choice leads to kinetic equations without complicated source terms and can be efficiently tackled by classical first-neighbour discrete velocity lattices. 
We refer interested readers to \citep{hosseini2024probing,strassle2025consistent} for an in-depth discussion.

Thus, the $f$-distribution function defines the fluid density and momentum, while the $g$-distribution defines the bulk energy \eqref{eq:rhoE},
\begin{align}
    &\int \{m,m\bm{v}\} f d\bm{v}= \{\rho, \rho\bm{u}\},\label{eq:fcons}\\
    &\int mg d\bm{v}= \rho E,\label{eq:gcons}
\end{align}
where $m$ is the mass of the particle.
The kinetic model is defined by the coupled kinetic equations,
\begin{align}
    &\partial_t f + \bm{v}\cdot\bm{\nabla} f = \mathcal{F},\label{eq:f_boltz}\\
    &\partial_t g + \bm{v}\cdot\bm{\nabla} g = \mathcal{G}\label{eq:g_boltz},
\end{align}
where the collision terms on the right-hand side are sought in the following form
\begin{align}
    & \mathcal{F} = \frac{1}{\tau}\left(f^{\rm eq} - f\right) + \frac{1}{\lambda}\left(f_\lambda^\star-f^{\rm eq}\right),\label{eq:f_coll}\\
    & \mathcal{G} = \frac{1}{\tau}\left(g^{\rm eq} - g\right) + \frac{1}{\lambda}\left(g_\lambda^\star-g^{\rm eq}\right).\label{eq:g_coll}
\end{align}
Here, the pair $\{f^{\rm eq}, g^{\rm eq}\}$ represents a local equilibrium attractor, while $\{f_\lambda^\star, g_\lambda^\star\}$ represents an intermediate quasi-equilibrium attractor. Furthermore, $\tau$ and $\lambda$ are the corresponding relaxation times, and 
we proceed to define the equilibrium and the quasi-equilibrium distribution functions.

To describe the equilibrium distribution functions, we introduce a \emph{reference temperature} parameter $\theta$, with dimensions of $RT$. This parameter fixes the width of the Maxwellian equilibria independent of the thermodynamic internal energy. The equilibrium associated with mass and momentum is then written as
\begin{equation}\label{eq:f_eq_cont}
    f^{\rm eq}(\rho,\bm{u},\theta)
    =
    \frac{\rho}{m{(2\pi \theta)}^{D/2}}
    \exp\left[-\frac{|\bm{v}-\bm{u}|^2}{2\theta}\right],
\end{equation}
where $\rho$ and the $D$ components of $\bm{u}$ are fixed by the local conservation constraints \eqref{eq:fcons}, while $\theta$ is specified independently. Thus, $f^{\rm eq}$ forms a $D+2$-parameter family.

The equilibrium distribution associated with the energy population is defined as
\begin{equation}\label{eq:geq_cont}
    g^{\rm eq}(\rho,\bm{u},T,\theta)
    =
    \left(
    \frac{\bm{v}^2}{2}
    +e(\rho,T)
    -\frac{D\theta}{2}
    \right)
    f^{\rm eq}(\rho,\bm{u},\theta),
\end{equation}
so that the thermodynamic temperature $T$ enters through the specific internal energy $e(\rho,T)$, whereas $\theta$ enters through the Maxwellian kernel. The equilibrium distributions \eqref{eq:f_eq_cont} and \eqref{eq:geq_cont} satisfy the conservation laws of mass, momentum, and bulk energy,
\begin{align}
    &\int \{m,m\bm{v}\} f^{\rm eq}\, d\bm{v}= \{\rho, \rho\bm{u}\},\label{eq:fcons_app}\\
    &\int mg^{\rm eq}\, d\bm{v}= \rho E.\label{eq:gcons_app}
\end{align}
Other relevant higher-order moments of the equilibrium distributions are presented in Table \ref{tab:moments_fg} in Appendix \ref{app:CE_kin}.

The quasi-equilibrium state $\{f_\lambda^\star, g_\lambda^\star\}$ is constructed using shifted values of the flow velocity, reference temperature, and thermodynamic temperature,
\begin{align}
    \bm{u}^\star_\lambda
    &= \bm{u}+\lambda\frac{\bm{F}}{\rho},
    \label{eq:sh_vel}\\
    \theta^\star_\lambda
    &= \theta + \lambda\alpha \theta
    \left(\bm{\nabla}\cdot\bm{u}\right),
    \label{eq:sh_temp} \\
    T_\lambda^\star
    &= T - \lambda^2\frac{\bm{F}\cdot\bm{F}}{2\rho^{2} c_v},
    \label{eq:sh_T}
\end{align}
where $\bm{F}$ is the Korteweg force \eqref{eq:FK}, and $\alpha$ is a non-dimensional parameter to be specified below. The quasi-equilibrium distribution $f_\lambda^\star$ therefore is a shifted Maxwellian,
\begin{equation}\label{eq:fstar_cont}
    f_\lambda^\star(\rho,\bm{u}_\lambda^\star,\theta_\lambda^\star)
    =
    f^{\rm eq}(\rho,\bm{u}_\lambda^\star,\theta_\lambda^\star).
\end{equation}
The corresponding quasi-equilibrium for the energy population is defined as
\begin{equation}\label{eq:gstar_cont}
    g^\star_\lambda(\rho,\bm{u}_\lambda^\star, T_\lambda^\star,\theta_\lambda^\star)
    =
    \left(
        \frac{\bm{v}^2}{2}
        + e(\rho, T_\lambda^\star)
        - \frac{D\theta_\lambda^\star}{2}
        + \frac{\bm{q}^{\rm c}_\lambda\cdot(\bm{v}-\bm{u})}{\rho \theta}
    \right)
    f^{\rm eq}(\rho,\bm{u}_\lambda^\star,\theta_\lambda^\star).
\end{equation}
The vector $\bm{q}^{\rm c}_\lambda$ appearing in the last term is defined by
\begin{equation}\label{eq:qc}
    \bm{q}^{\rm c}_\lambda = \lambda P\left(\bm{\nabla}h - \frac{k}{\mu}\bm{\nabla}T\right),
\end{equation}
where $h$ is the specific enthalpy:
\begin{equation}\label{eq:enthalpy}
h = e + \frac{P}{\rho}.
\end{equation}
Moments of the quasi-equilibrium distributions $f_\lambda^\star$ \eqref{eq:fstar_cont} and $g^\star_\lambda$ \eqref{eq:gstar_cont} are presented in Tables \ref{tab:moments_fg} and \ref{tab:moments_fgstar} in Appendix \ref{app:CE_kin}.

The construction above leaves two scalar parameters to be specified: the reference temperature $\theta$ and the non-dimensional coefficient $\alpha$ in \eqref{eq:sh_temp}. To specify these free parameters, we perform a Chapman--Enskog analysis of the hydrodynamic limit of the kinetic equations. Details are given in Appendix \ref{app:CE_kin}.
The analysis shows that, to recover the target hydrodynamic equations \eqref{eq:balance_density}, \eqref{eq:balance_momentum}, and \eqref{eq:balance_energy}, the reference temperature must be set as the thermodynamic flow work,
\begin{equation}
    \label{eq:theta}
    \theta=\frac{P}{\rho}.
\end{equation}
Recovery of the Navier--Stokes viscous stress tensor \eqref{eq:TNS} requires the relaxation time $\tau$ be related to the shear viscosity $\mu$ as
\begin{equation}
    \tau=\frac{\mu}{P},
\end{equation}
and the parameter $\alpha$ to have the form
\begin{equation}
    \label{eq:corr_bulk}
    \alpha=\left(\frac{D+2}{D}-\frac{\rho c_s^2}{P}-\frac{\eta}{\mu}\right).
\end{equation}
Note that the relaxation time $\lambda$ does not affect the hydrodynamic limit, thus remaining a free parameter which can be specified conveniently in the subsequent discretization.
We close this section with several explanatory comments.\\

\begin{enumerate}
\item In the ideal gas limit, $P\to \rho RT$, the reference temperature \eqref{eq:theta} becomes proportional to the thermodynamic temperature, $\theta\to RT$. For a generic non-ideal compressible fluid, the formulation of the reference temperature \eqref{eq:theta}, as a function of $P$ and $\rho$, rather than thermodynamic temperature $T$, was first proposed by \cite{reyhanian2020thermokinetic} in the context of the Particles-on-Demand method \citep{dorschner2018particles}.\\

\item A special case of kinetic models \eqref{eq:f_boltz} and \eqref{eq:g_boltz} was recently derived from the Boltzmann--Enskog--Vlasov kinetic theory by \cite{karlin2025practicalkineticmodelsdense} using a projection operator technique. Here, we introduced a more general kinetic model, combining the aforementioned approach of \cite{reyhanian2020thermokinetic} with the quasi-equilibrium representation of the non-local Korteweg's force.\\

\item In the absence of the term \eqref{eq:qc}, the nonequilibrium heat flux recovered in the hydrodynamic limit is proportional to the enthalpy gradient, $\bm{q}\propto\bm{\nabla}h$, as first observed by \cite{reyhanian2020thermokinetic}. In the ideal-gas limit this reduces to Fourier's law \eqref{eq:qF}, since $\bm{\nabla}h\propto\bm{\nabla}T$. For a general non-ideal fluid, however, $\bm{\nabla}h$ also contains density-gradient contributions, leading to a spurious non-Fourier component of the heat flux. The correction flux \eqref{eq:qc} compensates for this non-Fourier contribution, thereby recovering \eqref{eq:qF}.\\
    
\item {Setting $\alpha=0$ in \eqref{eq:sh_temp} recovers a fixed bulk viscosity 
$\eta=\mu\left(\frac{D+2}{D}-\frac{\rho c_s^2}{P}\right)$. 
Shifting the reference temperature with the parameter $\alpha$ \eqref{eq:corr_bulk} renders the bulk viscosity an independent, tunable, and positive-definite parameter.
}\\

\item {From the Chapman--Enskog analysis, the zeroth moment of $g_\lambda^\star$,
\begin{equation}\label{eq:total_e_expanded}
    \int m g_\lambda^\star d\bm{v}
    =
    \frac{1}{2}\rho\bm{u}^2
    + \lambda \bm{u}\cdot\bm{F}
    + \frac{\lambda^2}{2\rho}\bm{F}\cdot\bm{F}
    + \rho e(\rho,T_\lambda^\star),
\end{equation}
must differ from the corresponding moment of $g^{\rm eq}$ only by the term
$\lambda \bm{u}\cdot\bm{F}$. The additional kinetic energy introduced by the
shifted velocity $\bm{u}_\lambda^\star$ must therefore be compensated by a
commensurate reduction in the internal energy. Introducing
$\delta T = T_\lambda^\star - T$, we can expand
\begin{equation}\label{eq:e_expanded}
    e(\rho,T_\lambda^\star)
    =
    e(\rho,T)
    +
    \left.\frac{\partial e}{\partial T}\right|_{\rho}
    \delta T
    + \mathcal{O}\left(\delta T^2\right).
\end{equation}
To recover the correct moment up to order $\lambda^2$, one therefore requires
\begin{equation}
    \left.\frac{\partial e}{\partial T}\right|_{\rho}
    \delta T
    =
    -\frac{\lambda^2}{2\rho^2}\bm{F}\cdot\bm{F}.
\end{equation}
This yields the expression \eqref{eq:sh_T} for the shifted temperature.
Since $\left.\partial e/\partial T\right|_{\rho}=c_v>0$, the shifted temperature $T_\lambda^\star$ is systematically lower than $T$.}\\

\item 
    {The role of the quasi-equilibrium state can be clarified by considering the limit $\lambda\to0$, in which relaxation toward the shifted state becomes equivalent to introducing explicit kinetic source terms.}
    \begin{align}
&        \lim_{\lambda\to0}\frac{1}{\lambda}(f_\lambda^\star-f^{\rm eq})= \frac{1}{P}\bm{F}\cdot(\bm{v}-\bm{u})f^{\rm eq}+\alpha(\bm{\nabla}\cdot\bm{u})\left(\frac{{\rho(\bm{v}-\bm{u})}^2}{2P}-\frac{D}{2}\right)f^{\rm eq}, \label{eq:limitf}\\
\begin{split}\label{eq:limg}
 &       \lim_{\lambda\to0}\frac{1}{\lambda}(g_\lambda^\star-g^{\rm eq})=
        \frac{1}{P}\bm{F}\cdot(\bm{v}-\bm{u})\left(\frac{\bm{v}^2}{2} + e - \frac{DP}{2\rho}\right)f^{\rm eq}\\ +
        &\alpha(\bm{\nabla}\cdot\bm{u})\left(\left(\frac{\bm{v}^2}{2} + e - \frac{DP}{2\rho}\right)\left(\frac{{\rho(\bm{v}-\bm{u})}^2}{2P} - \frac{D}{2}\right)- \frac{DP}{2\rho}\right)f^{\rm eq}\\
    &    +{\left(\bm{\nabla}h - \frac{k}{\mu}\bm{\nabla}T\right)\cdot(\bm{v}-\bm{u})}f^{\rm eq}.
\end{split}
\end{align}
The first term in \eqref{eq:limitf} provides the Korteweg's force contribution to the momentum balance, while its counterpart, the first term in \eqref{eq:limg}, generates the corresponding work term in the bulk-energy balance equation \eqref{eq:balance_energy}. Furthermore, the second term in \eqref{eq:limitf} and its counterpart in \eqref{eq:limg} define the bulk viscosity contributions to the momentum and bulk-energy balance equations. Finally, the last term in \eqref{eq:limg} is the heat-flux correction required to recover Fourier's law, as discussed above.\\

\item {
The double-relaxation form of the kinetic equations \eqref{eq:f_boltz},\eqref{eq:g_boltz}, \eqref{eq:f_coll} and \eqref{eq:g_coll} has the advantage of admitting the standard lattice Boltzmann space--time discretization. Integration along characteristics combined with trapezoidal quadrature of the collision terms~\citep{he_1998_novel,ansumali_2007_quasi} yields an explicit, second-order accurate scheme (see details in Appendix \ref{app:int_chars}),
\begin{align}
%\begin{split}
    \label{eq:pre_LBM_f}
    f(\bm{x}+\bm{v}\delta t,t+\delta t)=&{f}+ 2\beta\left(f^{\rm eq} - {f}\right)+ {\frac{\delta t}{\lambda}}\left(1-\beta\right) \left(f_\lambda^\star- f^{\rm eq}\right),\\
    \label{eq:pre_LBM_g}
    g(\bm{x}+\bm{v}\delta t,t+\delta t)=&{g}+ 2\beta\left(g^{\rm eq} - {g}\right)+ {\frac{\delta t}{\lambda}}\left(1-\beta\right) \left(g_\lambda^\star - g^{\rm eq}\right).
%    \end{split}
\end{align}
Here all quantities on the right-hand side are evaluated at $(\bm{x}, t)$, $\delta t$ is the time step, and $\beta\in[0,1]$, is the transformed relaxation parameter
\begin{equation}\label{eq:beta_main}
    \beta=\frac{\delta t}{2\tau+\delta t}.
\end{equation}
As a consequence of the trapezoidal time integration, the transformed populations are related to the hydrodynamic fields $\rho$, $\bm{u}$ and $E$ as follows:
\begin{align}
\int m {f} d\bm{v} & = \rho,\label{eq:transform_rho}\\
\int m \bm{v} {f} d\bm{v} &= \rho\bm{u}-\frac{\delta t}{2}\bm{F},\label{eq:transform_u}\\
\int m {g} d\bm{v} &= \rho E - \frac{\delta t}{2} \bm{u}\cdot\bm{F}.\label{eq:transform_E}
\end{align}
}
\end{enumerate}

This concludes the presentation and analysis of the kinetic model for compressible non-ideal fluids.
The lattice Boltzmann realization follows.
}
%%%%%%%%%%%%%%%%%%%%%%%%%%%%%%%%%%%%%%%%%%%%%%%%%%%%%%%%%%%%%%%%%%%%%%%%%%%%%%%%%%%%%%%%%%%%%%%%%%%
\subsection{Lattice Boltzmann realization}\label{sec:LBrealization}
In the lattice Boltzmann realization of the kinetic model introduced above, we set $\lambda=\delta t$.
We consider the standard $D3Q27$ discrete velocity set $\bm{v}_i=c\bm{c}_i$ in $D=3$ dimensions, with $Q=27$ velocities,
\begin{equation}
    \label{eq:D3Q27c}
    \bm{c}_i=(c_{ix},c_{iy},c_{iz}),\ c_{i\alpha}\in\{-1,0,1\}.
\end{equation}
The $D3Q27$ lattice \eqref{eq:D3Q27c} is characterized by the lattice speed of sound,
\begin{equation}
    \label{eq:cs}
    \varsigma=\frac{1}{\sqrt{3}}c.
\end{equation}
In the following, we use lattice units by setting $c=1$.  The discrete velocity equations for the populations $f_i(\bm{x},t)$ and $g_i(\bm{x},t)$, $i=1,\dots, Q$, follow from \eqref{eq:pre_LBM_f} and \eqref{eq:pre_LBM_g}} as
{
\begin{align}
&f_i(\bm{x}+\bm{v}_i\delta t, t+\delta t)= f_i+ 2\beta\left(f_i^{\rm eq} - f_i\right)
+ \left(1-\beta\right) \left(f_i^{\star} - f_i^{\rm eq}\right),\label{eq:fLBGK}\\
&g_i(\bm{x}+\bm{v}_i\delta t, t+\delta t)= g_i+ 2\beta\left(g_i^{\rm eq} - g_i\right)
+ \left(1-\beta\right) \left(g_i^{\star} - g_i^{\rm eq}\right).\label{eq:gLBGK}
\end{align}
}
{It remains, then, to define the equilibrium populations $\{f_i^{\rm eq}, g_i^{\rm eq}\}$ and the shifted equilibrium populations $\{f_i^\star, g_i^\star\}$. To this end, we follow the product-form formalism \citep{karlin2010factorization} and introduce functions in two variables, $\xi_{\alpha}$ and $\zeta_{\alpha\alpha}$,}
%%%
    \begin{equation}
        \label{eq:phi}
 \Psi_{{i\alpha}}(\xi_{\alpha},\zeta_{\alpha\alpha})=1-c_{i\alpha}^2+ \frac{1}{2}\left[(3c_{i\alpha}^2-2)\zeta_{\alpha\alpha}+c_{i\alpha}\xi_{\alpha}\right],\ i=1,\dots,Q,\ \alpha=x,y,z.
    \end{equation}
The equilibrium populations $f_i^{\rm eq}$ are defined by setting the parameters {in the functions \eqref{eq:phi}} as follows: 
    \begin{align}
    &\xi_{\alpha}^{\rm eq}=u_{\alpha},\label{eq:def_xi_eq}\\
    &\zeta_{\alpha\alpha}^{\rm eq}={\theta}
    +u_{\alpha}^2,\label{eq:def_zeta_eq}
    \end{align}
{where $\theta$ is given by the thermodynamic flow work \eqref{eq:theta}.}
{With the definitions \eqref{eq:def_xi_eq} and \eqref{eq:def_zeta_eq} in the functions \eqref{eq:phi}, 
the local equilibrium populations are written in product-form,}
    \begin{equation}\label{eq:LBMeq}
         f_i^{\rm eq}= \rho\prod_{\alpha}\Psi_{{i\alpha}}\left(u_\alpha,{\theta}+u_{\alpha}^2\right).
    \end{equation}
{With $\lambda=\delta t$, the shifted flow velocity \eqref{eq:sh_vel} and shifted reference temperature  \eqref{eq:sh_temp} are
\begin{align}
    &\bm{u}^\star= \bm{u}+\delta t\left(\frac{\bm{F}}{\rho}\right),\label{eq:sh_vel_1}\\
    & \theta^\star = \theta + \delta t\alpha \theta
    \left(\bm{\nabla}\cdot\bm{u}\right).\label{eq:sh_temp_1}
\end{align}}
For the shifted-equilibrium populations $f_i^\star$, the parameters $\xi_\alpha$ and $\zeta_{\alpha\alpha}$ in the functions  \eqref{eq:phi} are set as follows:
\begin{align}
	&\xi_{\alpha}^{\star} = {u_{\alpha}^\star},\label{eq:xistar}	\\
	  &\zeta_{\alpha\alpha}^{\star} = 
      {\theta^\star}
      +{\left({u_{\alpha}^\star}\right)^2} + {\delta t}\Phi_{\alpha\alpha}.\label{eq:zetastar}
\end{align}
{
Compared with the continuous-velocity kinetic model of section~\ref{sec:kineticmodel}, the discrete-velocity formulation requires the additional correction term
\begin{equation}\label{eq:correction}
    \Phi_{\alpha\alpha} = -\frac{1}{\rho}\partial_{\alpha}\left(\rho u_{\alpha}^3 + {3\rho u_\alpha(\theta-\varsigma^2)} 
    \right).
\end{equation}
This is the standard correction necessary to restore Galilean invariance in the hydrodynamic limit on first-neighbour discrete-velocity lattices \citep{prasianakis2007lattice,li2012coupling,hosseini2020compressibility}.}
Combining \eqref{eq:xistar} and \eqref{eq:zetastar} with the same product-form construction utilizing \eqref{eq:phi}, the shifted-equilibrium populations may be written as
\begin{equation}\label{eq:LBMstar}
     f_i^\star=\rho\prod_{\alpha}\Psi_{{i\alpha}}\left({u_{\alpha}^\star},
     {\theta^\star}
     {
     + \left({u_{\alpha}^\star}\right)^2}+{\delta t}\Phi_{\alpha\alpha}\right),
\end{equation}
which completes the definition of the forcing term in \eqref{eq:fLBGK}. 

For the $g$-populations, we {follow the generating-function representation introduced in \citep{karlin2013consistent,saadat2021extended}.} The generating function is the bulk energy per unit mass,
\begin{equation}
    \label{eq:Egen}
    E(\rho,\bm{u},T) = e(\rho,T)+\frac{{u}^2}{2},
\end{equation}
consistent with \eqref{eq:rhoE}.
The corresponding equilibrium populations are constructed by repeated application of the operators
\begin{equation}\label{eq:Oa}
   { \mathcal{O}_\alpha (\theta)E = 
    \theta
        \dfrac{\partial E}{\partial u_\alpha}+ {u}_{\alpha} E,}
\end{equation}
whose dependence on the reference temperature is indicated explicitly.
The discrete equilibrium $g_i^{\rm eq}$ is defined by setting  $\xi_\alpha=\mathcal{O}_{\alpha}$ and $\zeta_{\alpha\alpha}=\mathcal{O}^2_{\alpha}$ in the functions \eqref{eq:phi} and interpreting the product-form as an operator acting on the generating function \eqref{eq:Egen},
    \begin{equation}\label{eq:LBMeqG}
         g_i^{\rm eq}\left(\rho,\bm{u},T,{\theta}\right)= \rho\prod_{\alpha}\Psi_{{i\alpha}}\left(\mathcal{O}_\alpha(\theta),[\mathcal{O}_\alpha(\theta)]^2\right)
         E(\rho,\bm{u},T).
    \end{equation}
{Shifted-equilibrium populations $g_i^\star$ are defined using the equilibrium product-form \eqref{eq:LBMeqG} evaluated at shifted values \eqref{eq:sh_vel_1} and \eqref{eq:sh_temp_1}, and adding a correction,}
\begin{equation}\label{eq:LBMgstar}
    g_i^\star = g_i^{\rm eq}\left(\rho,\bm{u}^\star, T^\star,
    {\theta^\star}
\right) 
    +
    \begin{cases}
    \dfrac{1}{2}\bm{c}_i\cdot\bm{q}^{\rm c}, & c_i^2=1,\\
    0, & \text{otherwise}.
\end{cases}
\end{equation}

where the non-equilibrium energy flux $\bm{q}^c$ and shifted temperature $T^\star$ are given by \eqref{eq:qc} and
\eqref{eq:sh_T}, respectively, with $\lambda=\delta t$:
\begin{align}\label{eq:sh_T_1}
    &T^\star = T - \delta t^2\left(\frac{\bm{F}\cdot\bm{F}}{2\rho^{{2}} c_v}\right),\\
    &{\bm{q}^{\rm c}=\delta t P\left(\bm{\nabla}h - \frac{k}{\mu}\bm{\nabla}T\right)}.\label{eq:qc_1}
\end{align}
Finally, the locally conserved fields entering the equilibrium and shifted-equilibrium populations, namely the density $\rho$, the momentum $\rho\bm{u}$ and the bulk energy $\rho E$, are defined from the zeroth- and first-order moments of the populations, {as in Eqs. \eqref{eq:transform_rho}, \eqref{eq:transform_u} and \eqref{eq:transform_E}, with sums over discrete velocities replacing velocity-space integrals},
\begin{align}
    &\rho=\sum_{i=1}^Q f_i,\label{eq:def_h}\\
    &\rho\bm{u}=\sum_{i=1}^Q \bm{c}_i f_i + \frac{\delta t}{2} \bm{F},
    \label{eq:def_u}\\
    &\rho E =\sum_{i=1}^Q g_i + \frac{\delta t}{2}\bm{u}\cdot\bm{F}.
\label{eq:def_E}
\end{align}
{Once the bulk energy $E$ is computed from Eq. \eqref{eq:def_E}, 
the thermodynamic temperature follows from the caloric equation obtained by integrating \eqref{eq:de_rho}.
For a van der Waals fluid, as the equation of state \eqref{eq:Pvdw} is linear in $T$, $c_v$ exhibits no density dependence. 
Integrating \eqref{eq:de_rho} therefore gives the explicit relation
\begin{equation}\label{eq:lbm_temp}
    T = \frac{E-\bm{u}^2/2+a\rho}{c_v}.
\end{equation}}
This explicit inversion is specific to equations of state with this simple caloric structure. For more general equations of state, such as Peng--Robinson, the temperature must instead be recovered at each node by solving the corresponding non-linear caloric relation.

Evaluating $\bm{F}$, $\Phi_{\alpha\alpha}$, and $\bm{q}^{\rm c}$ requires spatial derivatives of $\rho$, $T$, $h$, and $\bm{u}$. In the numerical applications below, all derivatives are computed using standard central second-order accurate finite differences, except for the first order derivative in \eqref{eq:correction}. This term is evaluated using an upwind-biased approximation, for instance along the $x-$axis,
\begin{equation}
    \Phi_{xx} = \frac{1+{\rm sgn}(u_x)}{2}\frac{\Lambda_x\lvert_x - \Lambda_x\lvert_{x-\delta x} }{\delta x} + \frac{1-{\rm sgn}(u_x)}{2}\frac{\Lambda_x\lvert_{x+\delta x} - \Lambda_x\lvert_{x} }{\delta x}
\end{equation}
where ${\rm sgn}(u_x)$ is the sign of $u_x$ and,
\begin{equation}
    \Lambda_x = \rho u_x^3 + {3\rho u_x(\theta-\varsigma^2)}.
\end{equation}
This upwind-biased approximation, while maintaining the formal order of accuracy of the solver has been shown to improve stability in higher Mach number simulations \citep{saadat_2021_extended,hosseini2020compressibility,renard2021improved}.

The overall structure of the proposed algorithm is shown in Fig.~\ref{Fig:FlowChart}.
\begin{figure*}
  \centering
  \begin{tikzpicture}[node distance=0.9cm and 2cm]

    % -------------------
    % Nodes in lanes
    % -------------------
    \node[block_m] (ini) {Initialize: $\{f_i,g_i\}=\{f_i^{\rm eq},g_i^{\rm eq}\}$};

    \node[block_m, below=of ini] (moments) {Compute $\rho$ \eqref{eq:def_h} and $\bm{F}$ \eqref{eq:FK}};

    % Larger node for correction
    \node[block_m, below=of moments, minimum width=8cm, text width=7.5cm] (correction)
    {Compute $\bm{u}$ \eqref{eq:def_u}, $E$ \eqref{eq:def_E}, $T$ \eqref{eq:lbm_temp}, $P$, $\Phi_{\alpha\alpha}$ \eqref{eq:correction}, and $\bm{q}^c$ \eqref{eq:qc_1}};

    \node[block_m, below=of correction] (fext) {Compute $\{f_i^\star, g_i^\star\}$: \eqref{eq:LBMstar} and \eqref{eq:LBMgstar}};

    \node[block_m, right=0.5cm of fext] (feq) {Compute $\{f_i^{\rm eq}, g_i^{\rm eq}\}$: \eqref{eq:LBMeq} and \eqref{eq:LBMeqG}};

    \node[block_m, below=2cm of fext] (CollStream) {Compute $\beta$ \eqref{eq:beta_main}, collide and stream \eqref{eq:fLBGK} and \eqref{eq:gLBGK}};

    \node[block_s, left=0.5cm of fext] (next) {$t = t + \delta t$};

    % -------------------
    % Main flow vertical arrows
    % -------------------
    \draw[->] (ini) -- (moments);
    \draw[->] (moments) -- (correction);
    \draw[->] (correction) -- (fext);
    \draw[->] (fext.south) -- (CollStream.north);

    % -------------------
    % Right lane arrows
    % -------------------
    \draw[->] (moments.east) -| (feq.north);

    \draw[->] (feq.south) |- (CollStream.east);

    % -------------------
    % Loop back
    % -------------------
    \draw[->] (CollStream.west) -| (next.south);
    \draw[->] (next.north) |- (moments.west);

  \end{tikzpicture}
  \caption{Overall structure of the proposed algorithm for the simulation of compressible non-ideal flows.}
  \label{Fig:FlowChart}
\end{figure*}
A detailed multi-scale analysis of the hydrodynamic limit is presented in Appendix~\ref{app:CE}, demonstrating that the present lattice Boltzmann model recovers the hydrodynamic equations \eqref{eq:balance_density}, \eqref{eq:balance_momentum} and \eqref{eq:balance_energy}. 
Numerical applications and validation of the model are presented in the next section.

\section{Numerical applications\label{sec:nums}}
\subsection{Consistency: Dispersion and dissipation of hydrodynamic modes}
\begin{figure}
	\centering
	\includegraphics[width=0.4\linewidth,keepaspectratio]{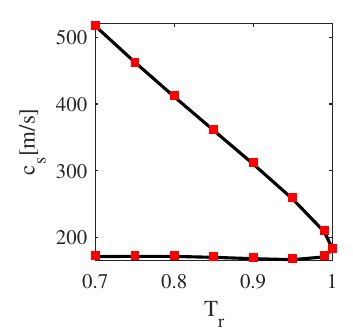}
	\caption{Speed of sound for nitrogen $\ce{N2}$ on the saturated liquid and vapour branches. Line: analytical solution from \eqref{eq:vdW_speed_of_sound}; markers: simulations.}
	\label{Fig:sound_speed}
\end{figure}

As a first step, we probe the dispersion and dissipation properties of the hydrodynamic shear, normal, and entropic modes in the limit of a resolved flow. These benchmarks will consider: (a) speed of sound, (b) shear wave dissipation, (c) shear stress, viscous heating, and entropic-mode dissipation, and (d) the normal mode dissipation rate. In all cases, and without loss of generality, we consider a van der Waals fluid fitted to the critical properties of nitrogen $\ce{N2}$, which are listed in Table \ref{tab:n2_critical}.
 {Below, the subscript $r$ refers to reduced variables normalized by their value at the critical point.}
\begin{table}
  \centering
  \caption{Critical properties of nitrogen\ce{N2}. The critical density used here comes from fitting critical temperature and pressure from~\citep{jacobsen1986thermodynamic} to the van der Waals equation of state.}
  \label{tab:n2_critical}
  \begin{tabular}{lcccc}
    \toprule
    
    Substance & $R/c_v$ & $P_c~[{\rm Pa}]$ & $\rho_c~[{\rm kg/m^3}]$ & $T_c~[{\rm K}]$ \\
    \midrule
    Nitrogen & 0.4 & $3.4\times10^6$ & 241.96 & 126.2 \\
    \bottomrule
  \end{tabular}
\end{table}
The setup for investigating the speed of sound consists of a one-dimensional domain of size $L_x = 0.1~[m]$ discretized with the spacing $\delta x=10 [\mu m]$. The initial conditions are
    \begin{align}
		  P(x) &= \begin{cases}
            P_0,  & x \leq L_x/2,\\
            P_0 + \delta P,  & x > L_x/2,
        \end{cases},\\
		u_x(x) &= u_{x0},\\
        T(x) &= T_0.
    \end{align}
Once the initial conditions are set, the system is left to evolve, resulting in two oppositely moving pressure fronts propagating at a speed that becomes constant after a short initial transition time. For sufficiently weak perturbations, this corresponds to the speed of sound in the system. A series of such cases with different initial conditions were run on both liquid and vapour branches of the saturation curve for {$T_0\in[88.34,126.2]~{\rm K}$ corresponding to $T_0/T_c\in[0.7,1]$} and compared to the analytical speed of sound \eqref{eq:speed_of_sound} for the van der Waals equation of state \eqref{eq:Pvdw},
\begin{equation}\label{eq:vdW_speed_of_sound}
    c_s = \sqrt{\frac{R T(1+R/c_v)}{{\left(1-b\rho\right)}^2} - 2 a\rho}.
\end{equation}
{Note that for a saturated vapour/liquid, fixing $T_0$ sets $\rho_0$ and $P_0$ via Maxwell's equal-area construction and the equation of state.} The results are shown in Figure \ref{Fig:sound_speed} and are in excellent agreement.

\begin{figure}
	\centering
	\includegraphics[width=0.4\linewidth,keepaspectratio]{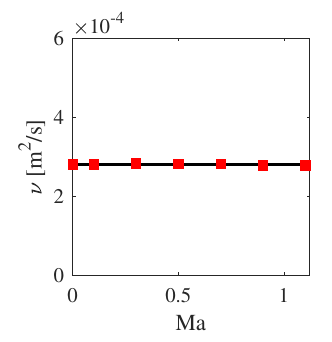}
	\caption{Kinematic viscosity as measured from shear wave decay simulations at different Mach numbers. Solid black line: analytical viscosity; square markers: viscosity measured from simulations.}
	\label{Fig:shear_dissipation}
\end{figure}

The next test is a measurement of the effective shear viscosity. For this, we set up a pseudo-one-dimensional, periodic domain of size $L_x=0.1~[m]$, discretized with $N_x\times N_y=100\times 1$ grid points. 
The initial conditions are set as follows:
\begin{eqnarray}
	u_y(x) &=& {\rm Ma}_y c_s(\rho_0, T_0) + \delta {\rm Ma}_y c_s(\rho_0, T_0) \sin{\left(\frac{2\pi x}{L_x}\right)},\\
	u_x(x) &=& 0,\\
    \rho(x) &=& \rho_0,\\
        T(x) &=& T_0.
\end{eqnarray}
The maximum amplitude of the perturbation, $u_y-{\rm Ma}_y c_s(\rho_0, T_0)$, in the domain is monitored throughout the simulation, and its evolution over time is fitted with an exponential decay function,
\begin{equation}
    u_{y}^{\rm max}(t) \propto \exp{\left(-{\frac{4\pi^2}{L_x^2}} \frac{\mu}{\rho} t\right)}.
\end{equation}
The viscosity measured from simulations is compared to that predicted from the multi-scale analysis. The results, shown in Fig.~\ref{Fig:shear_dissipation}, demonstrate excellent agreement and Galilean invariance of the measured viscosity.\\

\begin{figure}
	\centering
	\includegraphics[width=0.4\linewidth,keepaspectratio]{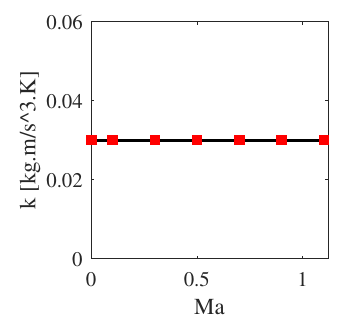}
	\caption{{Thermal conductivity as measured from simulations at different Mach numbers. Plain black line: analytical thermal conductivity; square markers: thermal conductivity measured from simulations.}}
	\label{Fig:entropic_dissipation}
\end{figure}

{Next we probe the thermal conductivity by monitoring the dissipation rate of temperature perturbations. The configuration consists of a pseudo-one-dimensional, periodic domain of size $L_x\times L_y=0.1~[m]\times 0.001~[m]$. Initial conditions are set as follows,
\begin{align}
	\rho(x) &= \rho_0 + \delta \rho \sin{\left(\frac{2\pi x}{L_x}\right)},\\
    P(x) &= P_0(\rho_0, T_0),\\
    % T(x) &= T_0 + \delta T,\\
    T(x) &= T(\rho(x), P_0),\\
	u_x(x) &= {\rm Ma}_x c_s(\rho_0, T_0),\\
    u_y(x) &= 0,
\end{align}
where \(T(\rho(x),P_0)\) is obtained from the equation of state.
We then monitor the maximum of the temperature in the domain $T^{\rm max}$ and extract the thermal conductivity by fitting the data to,
\begin{equation}
    \lvert T^{\rm max}-T_0\rvert \propto \exp\left(-\frac{4\pi^2}{L_x^2}\frac{k}{\rho_0 c_p(\rho_0,T_0)}t\right).
\end{equation}
The measured thermal conductivities are then compared to the imposed value. Results are shown in Fig.~\ref{Fig:entropic_dissipation} and are found to be in excellent agreement with the imposed value.}

\begin{figure}
	\centering
	\includegraphics[width=0.9\linewidth,keepaspectratio]{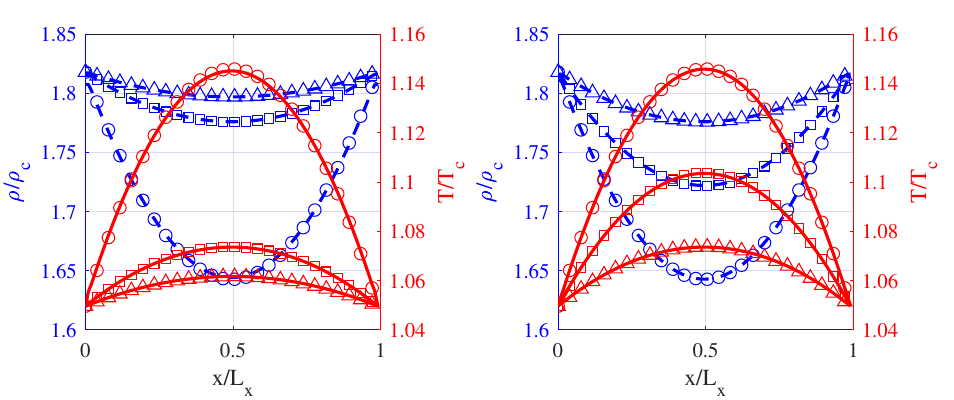}
	\caption{Left panel: Temperature and density distribution across the channel for the thermal Couette flow at different Prandtl numbers. Triangle, square and circular markers are analytical results for ${\rm Pr}\in\{0.6, 1.2, 4.9\}$ respectively. Solid and dashed lines are temperature and density profiles from simulations. Here ${\rm Ma}=0.8$ for all cases. Right panel: Temperature and density distribution for different Mach numbers. Triangle, square and circular markers are analytical results for ${\rm Ma}\in\{0.8, 1.2, 1.6\}$ respectively. Solid and dashed lines are temperature and density profiles from simulations. Here ${\rm Pr}=1.2$ for all cases.}
	\label{Fig:couette_vdW}
\end{figure}

\begin{figure}
	\centering
	\includegraphics[width=0.4\linewidth,keepaspectratio]{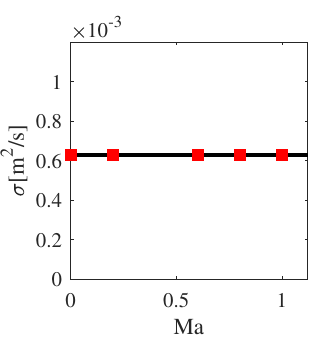}
	\caption{Normal dissipation rate $\sigma$ as measured from normal wave decay simulations at different Mach numbers. Plain black line: analytical dissipation rate, square markers: dissipation rate measured from simulations.}
	\label{Fig:normal_dissipation}
\end{figure}

To validate both viscous heating and the dissipation rate of the entropic modes, we next consider the two-dimensional thermal Couette flow. The case consists of a pseudo-one-dimensional domain of size $L_x$ with walls at $x=0$ and $x=L_x$ and periodic boundary conditions along the $y$-direction. The flow is subject to the following boundary conditions,
\begin{eqnarray}
	\{u_x, u_y, T\}(x=0) &=& \{0, 0, T_w\},\\
    \{u_x, u_y, T\}(x=L_x) &=& \{0, U_w, T_w\}.
\end{eqnarray}
The analytical steady-state solution to this configuration can readily be derived as
\begin{eqnarray}
	u_y(x) &=& U_w \frac{x}{L_x},\\
    T(x) &=& T_w + \frac{\mu U_w^2}{2k}\frac{x}{L_x}\left(1-\frac{x}{L_x}\right).
\end{eqnarray}
To validate the model, we consider a domain of size $L_x=1~[{\rm mm}]$ discretized with 100 grid points. Simulations were performed for ${\rm Pr}\in\{0.6, 1.2, 4.9\}$ and ${\rm Ma}\in\{0.8, 1.2, 1.6\}$. The results are displayed in Fig.~\ref{Fig:couette_vdW} and show excellent agreement with the analytical solutions.\\
Finally, we examine the dissipation rate of normal modes, i.e. acoustics. To do this, we set up a pseudo-one-dimensional domain of size $L_x$ with periodic boundary conditions in both the $x-$ and $y-$directions. Defining a uniform background state, $(\rho_0, T_0, P_0)$, we add a small perturbation to it at $t=0$,
\begin{align}
	u_x(x) &= {\rm Ma}_{x} c_s(\rho_0, T_0),\\
	u_y(x) &= 0,\\
    P(x) &= P_0 + \delta P \sin\left(\frac{2\pi x}{L_x}\right).
\end{align}
The density and temperature fields can be computed using isentropic relations for the van der Waals fluid \citep{kouremenos1988correlation,nederstigt2023generalised}, 
\begin{eqnarray}
	\frac{P_0}{\rho_0^{\gamma_{P\rho}^0}} &=& \frac{P}{\rho^{\gamma_{P\rho}}},\\
	\frac{T_0}{\rho_0^{\gamma_{T\rho}^0-1}} &=& \frac{T}{\rho^{\gamma_{T\rho}-1}},
\end{eqnarray}
where
\begin{eqnarray}
	\gamma_{P\rho} &=& \frac{\rho}{P}\frac{c_p}{c_v} \left(\frac{\partial P}{\partial \rho}\right)_T,\\
	\gamma_{T\rho} &=& \frac{1}{\rho c_v}\left(\frac{\partial P}{\partial T}\right)_\rho+1.
\end{eqnarray}
Here $c_p$ is the {specific heat at constant pressure},
\begin{equation}
    c_p = c_v + T{\left(\frac{\partial P}{\partial T}\right)}_v {\left(\frac{\partial v}{\partial T}\right)}_P,
\end{equation}
which for the van der Waals equation of state leads to
\begin{equation}\label{eq:cpvdW}
    c_p = c_v + \frac{R^2T}{RT-2a\rho{(1-b\rho)}^2}.
\end{equation}
In the ideal gas limit $P\to \rho R T$, both exponents reduce to $(\gamma_{P\rho},\gamma_{T\rho})\to {c_p}/{c_v}$. 
We leave the system to evolve over time and monitor the acoustic energy \citep{landau1987fluid},
\begin{equation}
    E_{\rm acoustic} = \frac{1}{2}\int\left[ \rho_0{|\bm{u}-\bm{u}_0|}^2 + \frac{{\rho'}^2 c_s^2(\rho_0, P_0)}{\rho_0} \right] dx,
\end{equation}
where $\rho'=\rho-\rho_0$. It can readily be shown that the decay for a propagating plane wave is proportional to,
\begin{equation}
    E_{\rm acoustic} \propto \exp{\left(-{\frac{4\pi^2}{L_x^2}} \sigma t\right)}.
\end{equation}
where~\citep{landau1987fluid}
{
\begin{equation}
    \sigma = \frac{2(D-1)}{D} \frac{\mu}{\rho}+\frac{\eta}{\rho} + \frac{k}{\rho c_p}\left(\frac{c_p}{c_v}-1\right).
\end{equation}}
Simulations were conducted for different initial velocities and effective dissipation rates measured. Results are shown in Fig.~\ref{Fig:normal_dissipation}.
The results show very good agreement with the analytical predictions.
\subsection{Multi-phase regime}

\subsubsection{Liquid-vapour co-existence}
Turning to the two-phase regime, we first probe the liquid--vapour co-existence densities as a validation of thermodynamic and mechanical consistency. Simulations are conducted in a pseudo-one-dimensional domain of size $L_x=0.4~[{\rm mm}]$ with periodic boundary conditions. The domain is filled with saturated vapor, with a column of saturated liquid in the center. Simulations are evolved until the density field converges.
\begin{figure}
	\centering
	\includegraphics[width=0.5\linewidth,keepaspectratio]{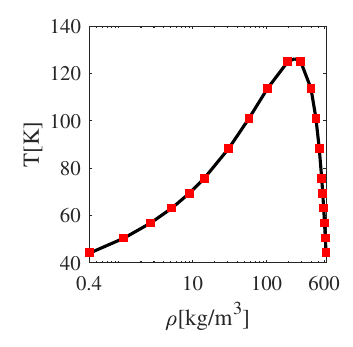}
	\caption{Liquid-vapour co-existence densities for nitrogen $\ce{N2}$. Line: Maxwell's equal area rule; symbols: simulation.}
	\label{Fig:coexistence_vdW_n2}
\end{figure}
Simulations were run for $T_r\in[0.3,0.99]$. As shown in Fig.~\ref{Fig:coexistence_vdW_n2}, numerical results agree closely with the reference values obtained using Maxwell's equal-area construction.
\subsubsection{Interface consistency and convergence}
\begin{figure}
	\centering
	\includegraphics[width=0.8\linewidth,keepaspectratio]{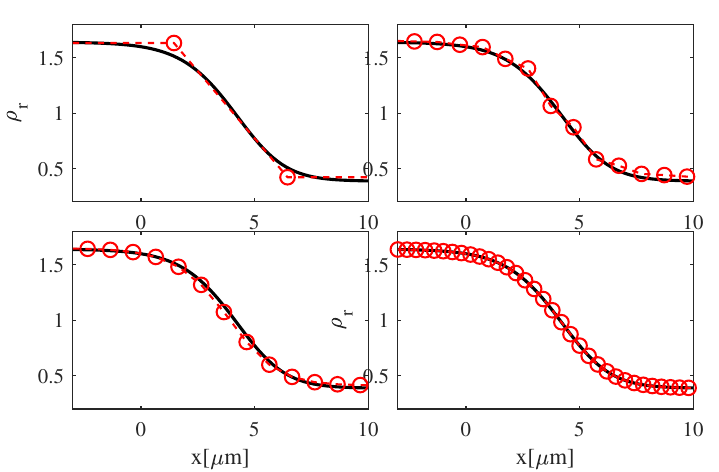}
	\caption{Liquid-vapour interface for nitrogen $\ce{N2}$ at $T_r=0.9$. Black lines are converged results from implicit finite-difference solver and red markers from LBM simulations. Top left panel: $\delta x=5~[\mu m]$ and top right panel: $\delta x=1~[\mu {\rm m}]$. Bottom left panel: $\delta x=0.5~[\mu {\rm m}]$. Bottom right panel: $\delta x=0.1~[\mu {\rm m}]$.}
	\label{Fig:interface_N2}
\end{figure}

\begin{figure}
	\centering
	\includegraphics[width=0.4\linewidth,keepaspectratio]{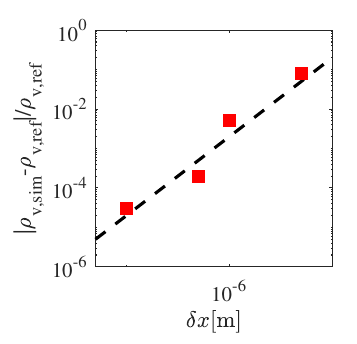}
	\caption{Convergence of the vapour-phase density for for nitrogen $\ce{N2}$ at $T_r=0.9$. Markers are results from simulations while the dashed line indicates second-order convergence.}
	\label{Fig:N2_convergence}
\end{figure}

To illustrate the consistency of the proposed solver, we repeat the nitrogen liquid--vapour co-existence density test described above, assessing the convergence of the interface with increasing resolution.
All physical parameters are unchanged: the nitrogen properties are those listed in
Table~\ref{tab:n2_critical}, the reduced temperature is $T_r=0.9$ with corresponding co-existence densities $(\rho^l,\rho^v)=(399.8,102.71)~[{\rm kg}/{\rm m}^3]$, and the
capillarity coefficient is set to
$\kappa=10^{-10}~{\rm m}^7\,{\rm kg}^{-1}\,{\rm s}^{-2}$.
The computational domain has a length of $L_x=0.5~{\rm mm}$.
{Simulations are conducted at the various resolutions detailed in Table \ref{tab:n2_interface}.}
\begin{table}
  \centering
  \caption{Grid properties for the nitrogen $\ce{N2}$ liquid--vapour interface simulations.}
  \label{tab:n2_interface}
  \begin{tabular}{lcc}
   \toprule
    Case & $\delta x$ & $\delta t$\\
   \midrule
    1 & $0.1~[\mu m]$ & $2.5\times10^{-11}[s]$\\
    2 & $0.5~[\mu m]$ & $1.25\times10^{-10}[s]$ \\
    3 & $1~[\mu m]$ & $2.5\times10^{-10}[s]$ \\
    4 & $5~[\mu m]$ & $1.25\times10^{-9}[s]$ \\
   \bottomrule
  \end{tabular}
\end{table}
The results obtained from simulations are compared to data from a high-resolution iterative finite-difference solver for
\begin{equation}
    \partial_x P = \kappa \rho \partial_x^3 \rho,
\end{equation}
and are shown in Fig.~\ref{Fig:interface_N2}.
The results demonstrate both excellent agreement with the reference solution and convergence under grid refinement. {To further quantify solver convergence, we compare the simulated vapour density with the reference value obtained from Maxwell's equal-area construction. The results, shown in Fig.~\ref{Fig:N2_convergence}, exhibit second-order convergence.}
\subsection{Compressible configurations}
\subsubsection{1-D non-ideal shock tubes}
\begin{figure}
	\centering
	\includegraphics[width=0.8\linewidth,keepaspectratio]{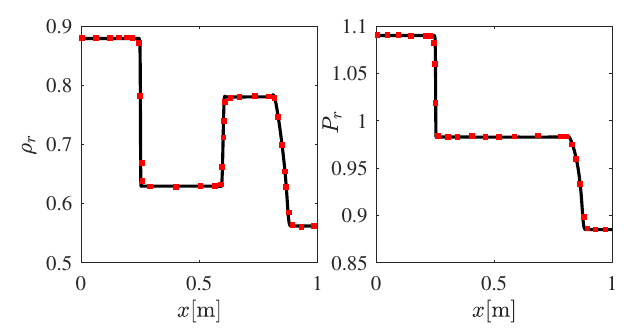}
	\caption{Reduced density and pressure fields for shock tube I at time $t=0.45 L_x\sqrt{\rho_c/P_c}$. Solid lines are reference data from \citep{guardone2002roe} and markers denote simulation results.}
	\label{Fig:shock_I}
\end{figure}

Until the early 1980s, shock-tube experiments were limited to gases exhibiting classical wave behaviour. \citet{borisov1983rarefaction} first reported a shock-tube experiment aimed at investigating non-classical wave phenomena in a dense gas, i.e. near the thermodynamic critical point. The nonlinear dynamics of gases are characterized, to a large extent, by the fundamental derivative of gas dynamics \citep{thompson1971fundamental}:
\begin{equation}\label{eq:Gamma}
    \Gamma = 1 + \frac{\rho}{c_s}\left(\frac{\partial c_s}{\partial \rho}\right)_s.
\end{equation}
For simple waves, $\Gamma$ represents the rate of change of the convected sound speed with respect to density. When $\Gamma>0$, the flow exhibits positive nonlinearity, i.e. disturbances steepen forward to form compression shocks. In contrast, when $\Gamma<0$, negative nonlinearity occurs and disturbances steepen backward, leading to expansion shocks. In regions of negative nonlinearity, gases display distinct non-classical phenomena.\\
{Following \citep{argrow1996computational}, we illustrate non-classical wave fields using three non-ideal shock-tube cases, each with flow regions wholly or partly in the regime of negative nonlinearity.}
In all three cases, simulations consist of a 1-D domain of size $L_x=1[{\rm m}]$, initially divided into left and right halves. The initial conditions set for each half are listed in Table~\ref{tab:shock_cases}.
\begin{table}
  \centering
  \caption{List of initial conditions for shock-tube cases.}
  \label{tab:shock_cases}
  \begin{tabular}{lccc}
    \toprule
    Case & $R/c_v$  & $(P_r,\rho_r)_{\rm left}$ & $(P_r,\rho_r)_{\rm right}$\\
    \midrule
    I & 0.0125 & (1.09,0.879) & (0.885,0.562)\\
    II & 0.329 & (1.6077,1.01) & (0.8957,0.594)\\
    III & 0.0125 & (3.00,1.818) & (0.575,0.275)\\
    \bottomrule
  \end{tabular}
\end{table}
In all cases, the grid size is set to $\delta x = 0.001~[m]$. Other simulation-specific parameters are listed in Table~\ref{tab:shock_cases_data}. {Note that the Courant–Friedrichs–Lewy (CFL) is here defined as the convective CFL,
\begin{equation}
    {\rm CFL} = \frac{\|\bm{u}\|}{\delta x/\delta t},
\end{equation}
where $\|\bm{u}\|=\sqrt{\bm{u}\cdot\bm{u}}$.}
\begin{table}
  \centering
  \caption{Numerical parameters for the shock-tube cases}
  \label{tab:shock_cases_data}
  \begin{tabular}{lcccc}
    \toprule
    Case & $\delta x$ & $\delta t$ & maximum CFL & maximum Ma\\
    \midrule
    I & $0.001[{\rm m}]$ & $8.3[\mu {\rm s}]$ & 0.4998 & 0.45\\
    II & $0.001[{\rm m}]$ & $4[\mu {\rm s}]$ & 0.4226 & 0.196\\
    III & $0.001[{\rm m}]$ & $2.86[\mu {\rm s}]$ & 0.4538 & 1.846\\
    \bottomrule
  \end{tabular}
\end{table}
In all cases, the initial discontinuity is located at half the length of the domain. The results for case I are shown and compared with the reference data in Fig. \ref{Fig:shock_I}.
This configuration is a typical example of non-classical gas dynamics: 
First, a compression front is observed moving from the high-pressure region to low-pressure region of the domain. In classical gas dynamics, one expects a sharp compression front which is not observed here. 
A second front, moving in the opposing direction, from the low-pressure region to the high-pressure region, known as a rarefaction front is also observed. Contrary to classical gas dynamic expectations, this front is sharp. Such non-classical wave fronts have recently been observed experimentally by \citet{colonna2026generation}.

\begin{figure}
	\centering
	\includegraphics[width=0.8\linewidth,keepaspectratio]{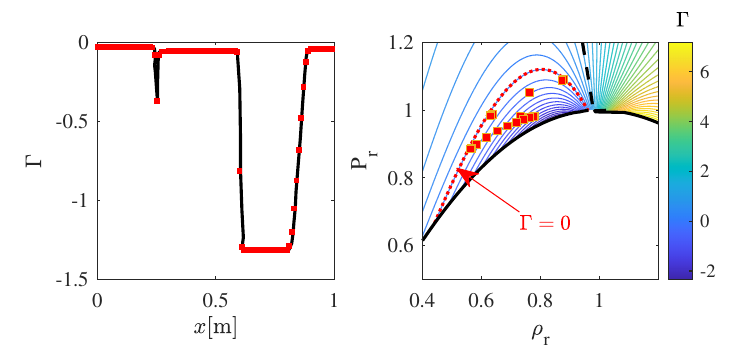}
	\caption{{(Left) Fundamental derivative $\Gamma$ distribution for shock tube I at $t=0.45 L_x\sqrt{\rho_c/P_c}$. Solid lines are reference data from \citep{guardone2002roe} and markers are from simulations. (Right) Fundamental derivative $\Gamma$ iso-contours in $P_r$--$\rho_r$ plane. Red square symbols represent the state of the shock tube shown in the left in the $P_r$--$\rho_r$ space. The solid black line is the co-existence curve.}}
	\label{Fig:shock_I_gamma}
\end{figure}
\begin{figure}
	\centering
	\includegraphics[width=0.8\linewidth,keepaspectratio]{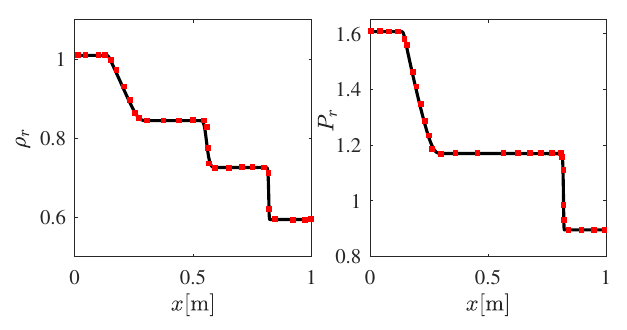}
	\caption{Reduced density and pressure fields for shock tube II at time $t=0.2 L_x\sqrt{\rho_c/P_c}$. Solid lines are reference data from \citep{guardone2002roe} and markers are from simulations.}
	\label{Fig:shock_II}
\end{figure}
\begin{figure}
	\centering
	\includegraphics[width=0.8\linewidth,keepaspectratio]{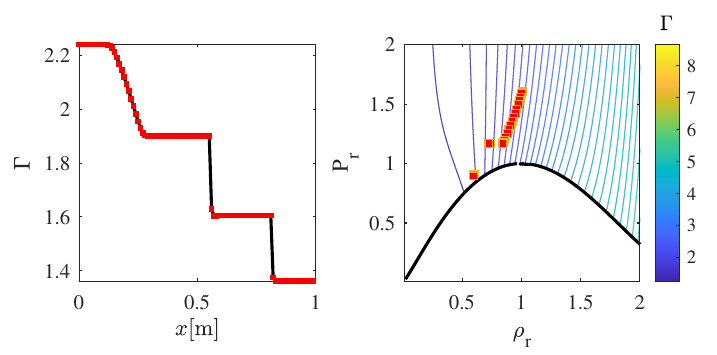}
	\caption{{(Left) Fundamental derivative $\Gamma$ distribution for shock tube II at $t=0.2 L_x\sqrt{\rho_c/P_c}$. Solid lines are reference data from \citep{guardone2002roe} and markers are from simulations. (Right) Fundamental derivative $\Gamma$ iso-contours in $P_r$--$\rho_r$ plane. red square symbols represent the state of the shock tube shown on the left in the $P_r$--$\rho_r$ space. The solid black line is the co-existence curve.}}
	\label{Fig:shock_II_gamma}
\end{figure}
\begin{figure}
	\centering
	\includegraphics[width=0.8\linewidth,keepaspectratio]{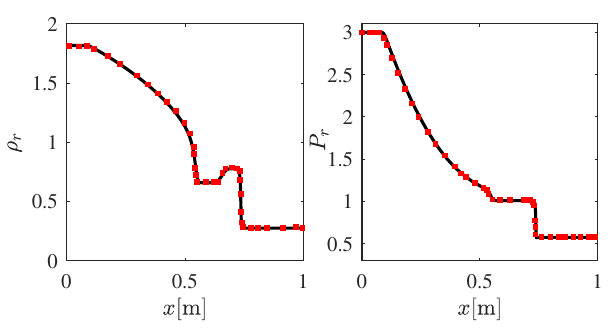}
	\caption{Reduced density and pressure fields for shock tube III at time $t=0.15 L_x\sqrt{\rho_c/P_c}$. Solid lines are reference data from \citep{guardone2002roe} and markers are from simulations.}
	\label{Fig:shock_III}
\end{figure}
\begin{figure}
	\centering
	\includegraphics[width=0.8\linewidth,keepaspectratio]{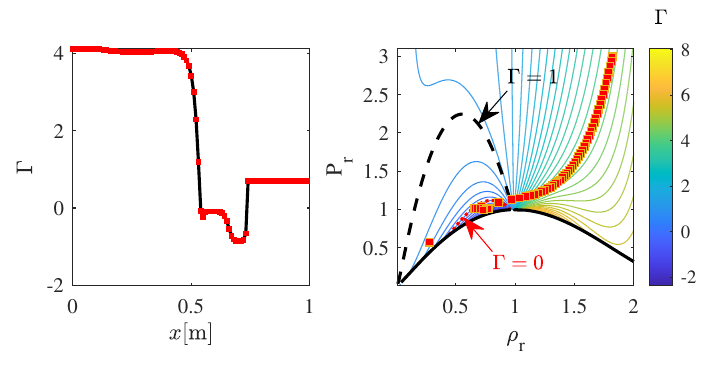}
	\caption{{(Left) Fundamental derivative $\Gamma$ distribution for shock tube III at $t=0.15 L_x\sqrt{\rho_c/P_c}$. Solid lines are reference data from \citep{guardone2002roe} and markers are from simulations. (Right) Fundamental derivative $\Gamma$ iso-contours in $P_r$--$\rho_r$ plane. Red square symbols represent the state of the shock tube shown in the left in the $P_r$--$\rho_r$ space. The solid black line is the co-existence curve.}}
	\label{Fig:shock_III_gamma}
\end{figure}

{These observations can be explained by looking into the fundamental derivative of gas dynamics \eqref{eq:Gamma},
which, in the case of a van der Waals fluid, can be computed explicitly as,
\begin{equation}
    \Gamma(P,\rho) = \frac{\left(R/c_v+1\right)\left(R/c_v+2\right)\frac{P+a\rho^2}{{(1/\rho-b)}^2} - 6a\rho^4}{2\left(R/c_v+1\right)\frac{P+a\rho^2}{{(1/\rho-b)}}-4a\rho^4}.
\end{equation}
It is observed in Fig.~\ref{Fig:shock_I_gamma} that all points in the domain at $t=0.45 L_x\sqrt{\rho_c/P_c}$ fall within the $\Gamma<0$ region. Physically, $\Gamma$ measures how the sound speed changes with compression or expansion along an isentrope. It directly controls the nonlinearity of acoustic waves. Consider first the compression front; along the front, both pressure and density decrease. When the $\Gamma<0$, the characteristic speeds decrease with increasing pressure (or density). Consequently, the local propagation speed is smaller in the higher-pressure region behind the front than in the lower-pressure region ahead of it. As a result, characteristics diverge across the compression region, causing the initially sharp front to spread. The compression wave therefore evolves into a compression fan rather than steepening into a discontinuity.
Next consider the rarefaction wave; Along this direction, pressure and density also decrease. For $\Gamma<0$, the characteristic speeds increase as pressure (or density) decreases. This leads to a convergence of characteristics, causing the rarefaction wave to steepen and the front to sharpen.} In Fig.~\ref{Fig:shock_I} one can clearly observe a rarefaction shock moving from right to left, i.e. low to high pressure. Additionally, one also observes a compression {fan} propagating into the low pressure region.\\
{In the second configuration, the fronts appear to follow the classical gas-dynamic behaviour.} As shown in Fig.~\ref{Fig:shock_II}, in agreement with reference data, both pressure and density fields show a compression front moving towards the low-pressure side and a rarefaction wave moving in the opposite direction.
{To further confirm the classical characteristics of this shock-tube case, we plot the distribution of $\Gamma$} at $t=0.2 L_x\sqrt{\rho_c/P_c}$ in Fig.~\ref{Fig:shock_II_gamma}.
In agreement with the reference data, $\Gamma$ remains positive throughout the domain.
As in the previous shock-tube configuration, the thermodynamic states sampled by the flow lie in the supercritical region and close to the saturation line. However, because of the considerably lower specific heat capacity compared to case I, no negative-$\Gamma$ region is encountered in the present case. This indicates that non-classical BZT effects are strongly influenced by the specific heat capacity and are therefore more likely to arise in fluids composed of larger, more complex molecules with higher specific heat capacities. \\
Finally, in the third shock-tube experiment both initial states remain in the classical region. Fig.~\ref{Fig:shock_III} shows the density and pressure profiles for this shock tube at $t=0.15 L_x\sqrt{\rho_c/P_c}$.
As in the previous configurations, a rarefaction front propagates leftward towards the dense region. Its structure, however, is non-classical: the front initially resembles a rarefaction fan, but sharpens near $x=0.5~[{\rm m}]$.  This behaviour can be explained using the fundamental derivative shown in Fig.~\ref{Fig:shock_III_gamma}. 
For $x\lesssim0.5~[{\rm m}]$, $\Gamma>1$ and the expansion behaves classically. Near $x=0.5~[{\rm m}]$, $\Gamma$ becomes negative; consequently the rarefaction wave steepens into a shock. A second sign change of $\Gamma$ occurs on the compression side, where the flow crosses back from the non-classical regime, associated with a compression fan, into the classical regime, where the compression front steepens into a shock.\\
All three configurations show excellent agreement with reference data and demonstrate that the model properly captures the behavior of a non-classical compressible gas.
\subsection{Shock--liquid-column interaction}
\begin{figure}
	\centering
	\includegraphics[width=0.6\linewidth,keepaspectratio]{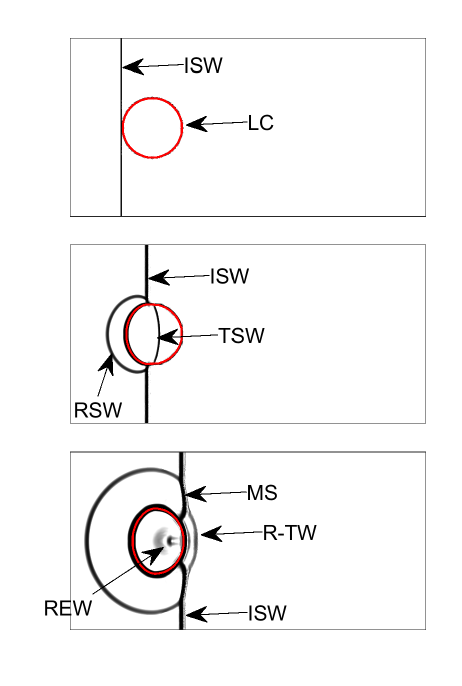}
	\caption{Schlieren images of shock--liquid-column interaction case at, from top to bottom, $t/t_0=0$, $t/t_0=0.3$ and $t/t_0=0.7$. Schlieren images are generated as $\phi = \exp\left(-a\frac{\|\bm{\nabla}\rho\|}{{\rm max}\left(\|\bm{\nabla}\rho\|\right)}\right)$ with $a=100$. The visualization follows~\citet{quirk1996dynamics,meng2015numerical}. Solid red lines indicate the liquid-column interface identified from a density level. ISW: incident shock wave; LC: liquid column; TSW: transmitted shock wave; RSW: reflected shock wave; MS: Mach stem; R-TW: retransmitted wave; REW: reflected expansion wave.}
	\label{Fig:shock_drop_ini_1o47}
\end{figure}

\begin{figure}
	\centering
	\includegraphics[width=0.9\linewidth,keepaspectratio]{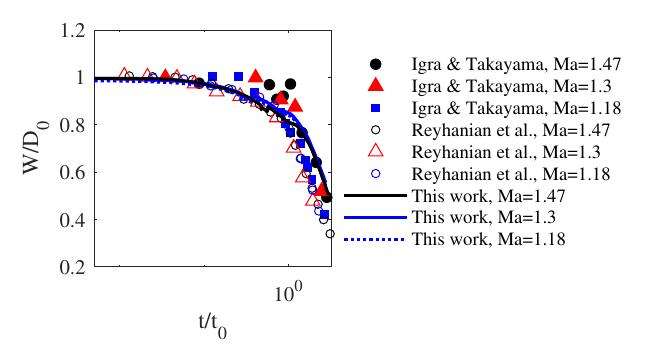}
	\caption{Evolution of the column width $W$ along $x$-axis over time for three different Mach numbers. Simulations: (solid black line) ${\rm Ma}_s=1.47$, (red dashed line) Ma=1.3 and (blue dotted line) ${\rm Ma}_s=1.18$. Experiments~\citep{igra2001study}: (Black filled circular markers) Ma=1.47, (Red filled triangle markers) ${\rm Ma}=1.3$ and (blue filled square markers) ${\rm Ma}=1.18$. Numerical results from \citep{reyhanian2020thermokinetic}: (black unfilled circular markers) ${\rm Ma}=1.47$, (red unfilled triangle markers) ${\rm Ma}=1.3$ and (blue unfilled square markers) ${\rm Ma}=1.18$}
	\label{Fig:shock_drop_time_1o47}
\end{figure}

\begin{figure}
	\centering
	\includegraphics[width=0.8\linewidth,keepaspectratio]{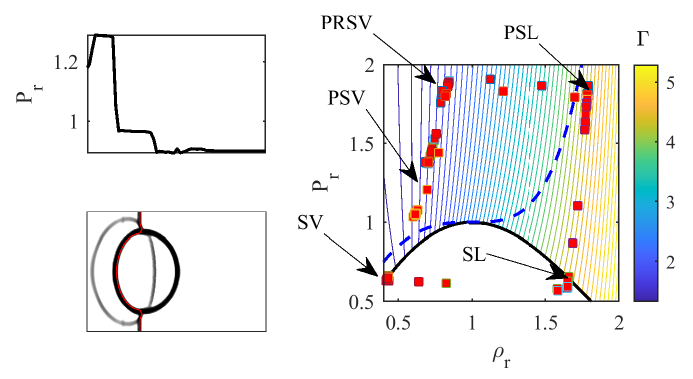}
	\caption{{Shock-liquid column interaction for ${\rm Ma}=1.47$ at $t/t_0=0.3$. (Bottom left) numerical Schlieren image. (Top left) reduced pressure distribution along the \(x\)-axis centreline. (Right) Fundamental derivative iso-contours in $P_r$--$\rho_r$ plane. Red square symbols represent the state of the domain. Blue dashed line the critical temperature isotherm. Solid black line is co-existence curve. PSV: Post-Shock Vapour, PRSV: Post-Reflected Shock Vapour, PSL: Post-Shock Liquid, SL: Saturated Liquid and SV: Saturated Vapour.}}
	\label{Fig:shock_drop_023state_147}
\end{figure}

In our final application, we showcase the suitability of the model for compressible regimes, by considering the case of a circular liquid column interacting with a planar shock-wave. The case consists of a two-dimensional domain of size $L_x\times L_y$, here resolved with $800\times800$ grid points, divided into subdomains via a shock positioned at $x_s$. On the right-hand side of the shock front, the pre-shock state $(\rho_1,T_1,u_{x,1})$ is set to that of a saturated vapour at $T_r=0.9$, following \citep{reyhanian2020thermokinetic}. The post-shock state to the left of the shock front, $(\rho_2,T_2,u_{x,2})$, is derived using the Rankine--Hugoniot conditions. Furthermore, a saturated liquid column of radius $R$ at $T_r=0.9$, resolved with $65$ grid points, is placed at $(x_c,y_c)$ in the pre-shock domain. Times are non-dimensionalized by the characteristic scale $t_0 = \frac{2R_0 c_s^v}{u_s c_s^l}\sqrt{\frac{\rho^l}{\rho^v}}$ where $u_s$ is the shock speed, $\rho^{l,v}$ are the liquid--vapour densities, and $c_s^{l,v}$ are the saturated liquid--vapour speeds of sound. The shock speed is defined via the shock Mach number ${\rm Ma_s}$ as $u_s = {\rm Ma}_s c_s^{v}$.\\
To further stabilize simulations, especially near sharp fronts, we use a non-linear numerical-viscosity scheme as devised in \citet{cook2004high,fiorina2005artificial}. This amounts to adding a numerical contribution to the transport coefficients as follows
\begin{equation}
    \{\mu,\eta,k\}^{\rm eff} = \{\mu,\eta,k\} + \{\mu,\eta,k\}^{\rm num}.
\end{equation}
The numerical contributions are defined as
\begin{equation}\label{eq:shock_cap}
    \{\mu,\eta\}^{\rm num} = \rho C_{\{\mu,\eta\}} \delta x^{r+1}\overline{\lvert \bm{\nabla}^{r-1}S\lvert},
\end{equation}
where
\begin{equation}
    S=\frac{1}{2}\sqrt{\left(\bm{\nabla}\bm{u}+\bm{\nabla}\bm{u}^\dagger\right):\left(\bm{\nabla}\bm{u}+\bm{\nabla}\bm{u}^\dagger\right)},
\end{equation}
and the overbar in \eqref{eq:shock_cap} indicates a Gaussian filter. Furthermore~\citep{fiorina2005artificial},
\begin{equation}\label{eq:shock_cap_k}
    k^{\rm num} = \rho c_s C_k \delta x^{r+1}\overline{\lvert \bm{\nabla}^{r-1}\lvert\bm{\nabla}s\rvert\rvert},
\end{equation}
with the entropy gradient obtained from \eqref{eq:ds_vdW} as\begin{equation}
{\bm{\nabla}s=\frac{c_v}{T}\bm{\nabla}T-\frac{R}{\rho(1-b\rho)}\bm{\nabla}\rho.}
\end{equation}
Here, to minimize the numerical dissipation we have set $r=5$. The flow evolution, represented by a Schlieren image for ${\rm Ma}_s=1.47$ is shown in Fig.~\ref{Fig:shock_drop_ini_1o47}.
The wave structures that arise during the initial stages of the shock--liquid-column interaction are commonly used to validate numerical schemes. In the present study, representative wave patterns are extracted and illustrated in Figure~\ref{Fig:shock_drop_ini_1o47}. Only the early-stage interaction between a planar shock wave and a cylindrical liquid column is considered here. As the incident shock wave travels from left to right across the liquid column, both a transmitted wave and a reflected shock wave are generated. The reflected shock wave propagates upstream into the surrounding vapour, while the transmitted wave moves downstream within the liquid column. Notably, the transmitted shock wave moves faster than the incident shock wave, as the speed of sound in the liquid is larger than that in the vapour phase. Upon reaching the downstream interface of the column, the transmitted wave re-emerges into the downstream vapour. Simultaneously, expansion waves reflect repeatedly within the liquid column. At the upper lateral edge of the liquid column, the incident shock wave, the Mach stem, and the slip line intersect to form a triple point. These wave structures along with the two-phase interface represent characteristic features of early-stage shock-liquid-column interaction. They appear as discontinuities of varying intensity, posing significant challenges for numerical modelling. The liquid column subsequently flattens in the flow direction and expanding in the transverse direction. As further quantitative validation, Figure \ref{Fig:shock_drop_time_1o47} shows the evolution of the width of the column $W$ along the centreline, compared with experiments and numerical simulations as reported in \citep{reyhanian2020thermokinetic} for three Mach numbers. 
The results are in good agreement with both the experiments and simulations showing that the deformation of the liquid column was accurately captured by the proposed scheme.\\
In addition to the flow features discussed above, we examine the thermodynamic states sampled by the solution. Unlike previously documented configurations, the present case evolves near the critical point. Figure~\ref{Fig:shock_drop_023state_147} shows the thermodynamic state at $t/t_0=0.3$.
Along the centreline, from left to right, the fluid first occupies the post-shock vapour state labelled PSV, which lies above both the critical pressure and the critical-temperature isotherm and is therefore supercritical. Behind the reflected shock, the pressure and temperature increase sharply, with the pressure reaching nearly twice the critical value. Across the liquid-column interface, the pressure remains approximately constant while the temperature decreases. This brings the fluid below the critical-temperature isotherm to the state labelled PSL. Inside the liquid column, the shock reduces both pressure and temperature, driving the fluid toward the saturated-liquid state.
Overall, the shock interaction heats the initially saturated liquid column to temperatures close to, but slightly below, the critical temperature, while the surrounding vapour is driven well above it. The resulting configuration is a supercritical vapour surrounding in contact with a marginally subcritical liquid column.
\section{\label{sec:level4}Conclusion} 
{The development of numerical methods for compressible non-ideal fluid dynamics remains comparatively underexplored, despite the growing relevance of such flows in modern energy systems. In this work, we have proposed a kinetic framework and its lattice Boltzmann realization specifically designed to address this gap.

At the core of the approach lies a thermodynamically consistent kinetic model for dense fluids, based on two BGK-type collision operators with Gaussian attractors defined by a local reference state and a shifted equilibrium. This construction enables the incorporation of the full non-ideal thermodynamic pressure as the local reference attractor while accounting for the inherently non-local nature of intermolecular interactions. In contrast to formulations relying on a single BGK operator carrying the full pressure—which may lead to non-physical behaviour such as non-positive bulk viscosity and density-gradient-dependent energy diffusion in the heat flux—the present model ensures positive-definite transport properties and recovers the target macroscopic balance equations under appropriate scaling.

We then developed a consistent lattice Boltzmann discretization yielding a second-order accurate numerical scheme. The resulting model was validated across a range of canonical configurations. In particular, shock-tube simulations revealed the emergence of non-classical wave dynamics, including rarefaction shocks and mixed rarefaction fan–shock structures associated with regions of negative fundamental derivative. Simulations of a shock wave interacting with a liquid column near critical conditions further demonstrated the ability of the model to capture strong thermodynamic effects, such as shock-induced transitions toward supercritical states. In the shock--liquid-column interaction case, the initially saturated liquid column experiences a strong temperature rise following shock impact. The liquid-column temperature approaches the critical value, while the surrounding vapour exceeds it, resulting in a marginally subcritical liquid column, surrounded by a  supercritical vapour.

In the hydrodynamic limit, corresponding to the long-wavelength limit, both the kinetic formulation and its lattice Boltzmann discretization are stable for arbitrary parameter choices. For finite wavenumbers, which are present in most practical simulations and particularly in strongly compressible configurations such as shock tubes, a practical stability condition commonly observed in third-order quadrature-based schemes is $\theta < \delta x^2/(3\delta t^2)$. This constraint appears largely independent of the particular collision operator employed. In addition, regimes where $\beta$ approaches unity may lead to numerical oscillations due to the increasing influence of higher-order kinetic moments. In principle, such effects can be mitigated using multiple-relaxation-time collision models, which allow improved control over higher-order moments such as the entropic multiple-relaxation-time collision operator~\citep{karlin2014gibbs}. This will be addressed in upcoming publications.

Overall, the results highlight the capability of the proposed framework to reproduce key non-ideal compressible phenomena within an efficient kinetic formulation. The thermodynamic consistency and robustness of the model make it a promising tool for the investigation of complex flows involving strong departures from ideal-gas behaviour, including regimes relevant to supercritical technologies and phase-transition-driven processes such as flash boiling.}

\section*{Acknowledgment}
This work was supported by European Research Council (ERC) Advanced Grant No. 834763-PonD and by the Swiss National Science Foundation (SNSF) Grants 200021-228065 and 200021-236715. Computational resources at the Swiss National Super Computing Center (CSCS) were provided under Grants No. s1286, sm101 and s1327. Authors would like to thank Patrick Jenny for his support and fruitful discussions.

\section*{Declaration of interests}
The authors report that they do not have a conflict of interest.

\section*{Data Availability Statement}
The data that support the findings of this study are available from the corresponding author upon request.

\appendix

\section{Multi-scale analysis of kinetic model\label{app:CE_kin}}
{
To derive the hydrodynamic limit of the kinetic model, we consider the time-evolution equations
\begin{equation}
	\partial_t \{f, g\} + \bm{v}\cdot\bm{\nabla} \{f, g\} 
    = 
    \frac{1}{\tau} \left( \{f^{\rm eq}, g^{\rm eq}\} - \{f, g\}\right) 
    + \frac{1}{\lambda} \left( \{f_\lambda^{\star}, g_\lambda^{\star}\} - \{f^{\rm eq}, g^{\rm eq}\}\right).
\end{equation}

For clarity, we restrict attention to the limit $\lambda \rightarrow 0$. In this limit, the shifted equilibrium operators reduce to
\begin{equation}
    \lim_{\lambda\rightarrow0} \frac{1}{\lambda} \left( \{f_\lambda^{\star}, g_\lambda^{\star}\} - \{f^{\rm eq}, g^{\rm eq}\}\right) = \{\mathcal{F}^\star, \mathcal{G}^\star\},
\end{equation}
where
\begin{align}
\mathcal{F}^\star &= \left[\frac{\bm{F}\cdot(\bm{v}-\bm{u})}{P}+\alpha(\bm{\nabla}\cdot\bm{u})\left(\frac{\rho(\bm{v}-\bm{u})^2}{2P}-\frac{D}{2}\right)\right]f^{\rm eq}, \label{eq:limf}\\
\mathcal{G}^\star &= \left(\frac{\bm{v}^2}{2}+e(\rho,T)-\frac{DP}{2\rho}\right)\mathcal{F}^\star + \frac{(\bm{v}-\bm{u})\cdot\bm{q}^c}{P}f^{\rm eq},
\end{align}
with
\begin{equation}
    \bm{q}^c = P\left(\bm{\nabla}h-\frac{k}{\mu}\bm{\nabla}T\right).
\end{equation}

We next introduce a non-dimensional formulation based on the characteristic scales
\begin{itemize}
    \item flow velocity $\mathcal{U}$,
    \item length scale $\mathcal{L}$,
    \item time scale $\mathcal{T}=\mathcal{L}/\mathcal{U}$,
    \item reference density $\bar{\rho}$,
    \item speed of sound $c_s$ (see equation~\eqref{eq:speed_of_sound}),
    \item interface thickness $\delta$.
\end{itemize}

The corresponding dimensionless variables are defined as
\begin{itemize}
    \item $t=\mathcal{T}t'$,
    \item $\bm{x}=\mathcal{L}\bm{x}'$,
    \item $\bm{u}=\mathcal{U}\bm{u}'$,
    \item $\bm{v}=c_s\bm{v}'$,
    \item $\rho=\bar{\rho}\rho'$,
    \item $f=\bar{\rho}c_s^{-3}f'$,
    \item $g=\bar{\rho}c_s^{-1}g'$.
\end{itemize}

This introduces the following dimensionless groups:
\begin{itemize}
    \item Knudsen number ${\rm Kn}=\tau c_s/\mathcal{L}$,
    \item Mach number ${\rm Ma}=\mathcal{U}/c_s$,
    \item Cahn number ${\rm Ca}=\delta/\mathcal{L}$.
\end{itemize}

The dimensionless evolution equations then read
\begin{multline}
	\partial'_t f' + \frac{1}{\rm Ma}\bm{v}'\cdot\bm{\nabla}' f' 
    = 
    \frac{1}{{\rm Ma}{\rm Kn}}\left( (f^{\rm eq})' - f'\right) 
    \\ + \left[\frac{{\rm Ca}^2}{\rm Ma}\frac{\bm{F}'\cdot(\bm{v}'-\bm{u}')}{P'}+\frac{1}{\rm Ma}\alpha'(\bm{\nabla}'\cdot\bm{u}')\left(\frac{\rho'(\bm{v}'-\bm{u}')^2}{2P'}-\frac{D}{2}\right)\right](f^{\rm eq})',
\end{multline}
and
\begin{multline}
	\partial'_t g' + \frac{1}{\rm Ma}\bm{v}'\cdot\bm{\nabla}' g' 
    = 
    \frac{1}{{\rm Ma}{\rm Kn}}\left( (g^{\rm eq})' - g'\right) 
    + \frac{1}{\rm Ma}\frac{(\bm{v}'-\bm{u}')\cdot{\bm{q}^c}'}{P'}(f^{\rm eq})'
    \\+ \left[\frac{{\rm Ca}^2}{\rm Ma}\frac{\bm{F}'\cdot(\bm{v}'-\bm{u}')}{P'}+\frac{1}{\rm Ma}\alpha'(\bm{\nabla}'\cdot\bm{u}')\left(\frac{\rho'(\bm{v}'-\bm{u}')^2}{2P'}-\frac{D}{2}\right)\right](g^{\rm eq})'.
\end{multline}

Finally, we adopt the scaling assumptions
\begin{itemize}
\item acoustic scaling ${\rm Ma}\sim O(1)$,
\item hydrodynamic scaling ${\rm Kn}\sim \epsilon$,
\item capillary scaling ${\rm Ca}\sim O(1)$.
\end{itemize}

Under these assumptions, the system reduces to
\begin{equation}
    \epsilon \left( \partial_t \{f, g\} + \bm{v}\cdot\bm{\nabla} \{f, g\} \right) = \left( \{f^{\rm eq}, g^{\rm eq}\} - \{f, g\} \right) 
    + \{\mathcal{F}^{\star}, \mathcal{G}^{\star}\},
\end{equation}
where primes have been omitted for clarity.\\
To probe the hydrodynamic limit, $\epsilon\rightarrow0$, we introduce a multi-scale  expansion in terms of the smallness parameter $\epsilon$,
\begin{align}
    \{f, g\} &= \{f^{(0)}, g^{(0)}\} + \epsilon \{f^{(1)}, g^{(1)}\} + \epsilon^2 \{f^{(2)}, g^{(2)}\} 
    + \mathcal{O}(\epsilon^3),\\
    \{\mathcal{F}^{\star}, \mathcal{G}^{\star}\} &=  \epsilon \{\mathcal{F}^{\star(1)}, \mathcal{G}^{\star(1)}\},
\end{align}
and,
\begin{equation}
    \partial_t = \partial_t^{(1)} + \epsilon \partial_t^{(2)} + + \mathcal{O}(\epsilon^3),
\end{equation}
into the dimensional equations and separate in terms of different orders of $\epsilon$,
\begin{align}
    \epsilon^0:\ \ &%
        0 = \frac{1}{\tau} \left( \{f^{\rm eq}, g^{\rm eq} - \{f^{(0)} , g^{(0)}\}\right)
        ,\label{Eq:CE_Eq_orders_0}
\\
    \epsilon^1:\ \ &%
        \partial_t^{(1)} 
        \{f^{(0)}, g^{(0)}\}  + \bm{v}\cdot\bm{\nabla} \{f^{(0)}, g^{(0)}\} = 
        -\frac{1}{\tau} \{f^{(1)} , g^{(1)}\} + \{\mathcal{F}^{\star(1)}, \mathcal{G}^{\star(1)}\}
        ,\label{Eq:CE_Eq_orders_1}
\\
    \epsilon^2:\ \ &%
        \partial_t^{(1)} 
        \{f^{(1)}, g^{(1)}\}  + \bm{v}\cdot\bm{\nabla} \{f^{(1)}, g^{(1)}\} + \partial_t^{(2)}\{f^{(0)}, g^{(0)}\} = 
        -\frac{1}{\tau} \{f^{(2)} , g^{(2)}\} + \{\mathcal{F}^{\star(2)}, \mathcal{G}^{\star(2)}\}
        .\label{Eq:CE_Eq_orders_2}
\end{align}
From order $\epsilon^0$ it directly follows that
\begin{equation}
    \{f^{(0)}, g^{(0)}\}=\{f^{\rm eq}, g^{\rm eq}\}.
\end{equation}
Before moving on to the next orders, for the sake of readability we have listed the moments of $f^{\rm eq}$, $g^{\rm eq}$, $\mathcal{F}^\star$ and $\mathcal{G}^\star$ in tables ~\ref{tab:moments_fg} and \ref{tab:moments_fgstar}.\\
\begin{table}
\centering
\small
\setlength{\tabcolsep}{4pt}
\renewcommand{\arraystretch}{1.2}
\begin{tabular}{ccc} % no vertical bars
\toprule
\textbf{Moment} & $f^{\rm eq}$ & $g^{\rm eq}$ \\
\midrule
1 & $\rho$ & $\rho E$ \\
$v_\alpha$ & $\rho u_\alpha$ & $\rho (E+P/\rho) u_\alpha$ \\
$v_\alpha v_\beta$ & 
$\rho u_\alpha u_\beta + P \delta_{\alpha\beta}$ & 
$\rho u_\alpha u_\beta(E+2P/\rho)+ P(E+P/\rho)\delta_{\alpha\beta}$\\
$v_\alpha v_\beta v_\gamma$ &
$\rho u_\alpha u_\beta u_\gamma+ P ( u_\alpha \delta_{\beta\gamma} + u_\beta \delta_{\alpha\gamma} + u_\gamma \delta_{\alpha\beta})$ &
-- \\
\bottomrule
\end{tabular}
\caption{Moments of $f^{\rm eq}$ and $g^{\rm eq}$.}
\label{tab:moments_fg}
\end{table}

\begin{table}
\centering
\small
\setlength{\tabcolsep}{4pt}
\renewcommand{\arraystretch}{1.2}
\begin{tabular}{ccc} % no vertical bars
\toprule
\textbf{Moment} & $\mathcal{F}^\star$ & $\mathcal{G}^\star$ \\
\midrule
1 & $0$ & $\bm{u}\cdot\bm{F}$ \\
$v_\alpha$ & $F_\alpha$ & 
$u_\alpha(\bm{u}\cdot\bm{F})+ q_{\alpha}^c$\\
$v_\alpha v_\beta$ & 
$u_\alpha F_\beta + u_\beta F_\alpha+P\alpha(\bm{\nabla}\cdot\bm{u})$ & 
$(u_\alpha u_\beta+\delta_{\alpha\beta}\frac{P}{\rho})(\bm{u}\cdot\bm{F})+\frac{P}{\rho}(u_\alpha (F_\beta+\frac{\rho q^c_\beta}{P})+u_\beta (F_\alpha+\frac{\rho q^c_\alpha}{P}))$\\
\bottomrule
\end{tabular}
\caption{Moments of $\mathcal{F}^\star$ and $\mathcal{G}^\star$.}
\label{tab:moments_fgstar}
\end{table}
Next, going to order $\epsilon$ and computing the moments $\int\{f, \bm{v} f, g\}d\bm{v}$ of the Chapman--Enskog-expanded equations, using solvability conditions,
\begin{align}
    &\int m\{1,\bm{v}\} f^{(k)} d\bm{v}= 0,\,\forall k>0,\\
    &\int mg^{(k)} d\bm{v}= 0,\,\forall k>0,
\end{align}
and the moments listed in tables~\ref{tab:moments_fg} and \ref{tab:moments_fgstar} we obtain,
\begin{gather}
	    \partial_t^{(1)}\rho + \bm{\nabla}\cdot \rho\bm{u}  = 0
        ,
        \label{eq:Eulerlevel_continuity1}
\\
	    \partial_t^{(1)}\left( \rho \bm{u} \right) + \bm{\nabla}\cdot \left( \rho \bm{u}\otimes\bm{u} + P\bm{I} \right) = -\bm{\nabla}\cdot\bm{T}_K
        ,
        \label{eq:Eulerlevel_momentum1}
\\
	    \partial_t^{(1)}\left( \rho E \right) + \bm{\nabla}\cdot \left( \rho E\bm{u} + P\bm{u}\right) = -\bm{u}\cdot\bm{\nabla}\cdot\bm{T}_K
        .
        \label{eq:Eulerlevel_energy1}
\end{gather}
The last equation, Eq.~\eqref{eq:Eulerlevel_energy1} can be transformed into a balance equation for internal energy, using ,
\begin{equation}
    \partial_t^{(1)} \mathcal{K} + \bm{\nabla}\cdot(\bm{u} \mathcal{K}) + \bm{u}\cdot\bm{\nabla}P + \bm{u}\cdot\bm{\nabla}\cdot\bm{T}_K = 0,
\end{equation}
as,
\begin{equation}
    \partial_t^{(1)} \rho e + \bm{\nabla}\cdot\rho \bm{u} e + P\bm{\nabla}\cdot\bm{u}  = 0.
\end{equation}
Furthermore, using
\begin{equation}
    de = c_v dT - \left(T \left(\frac{\partial P}{\partial T}\right)_\rho- P\right)\frac{d\rho}{\rho^2},
\end{equation}
and Eq.~\eqref{eq:Eulerlevel_continuity1} a balance equation for temperature can be derived as,
\begin{equation}
    \partial_t^{(1)}T + \bm{u}\cdot\bm{\nabla}T + \frac{T}{\rho c_v}\left(\frac{\partial P}{\partial T}\right)_\rho\bm{\nabla}\cdot\bm{u} = 0.
\end{equation}
Finally, using
\begin{equation}
    dP = \left(\frac{\partial P}{\partial \rho}\right)_T d\rho + \left(\frac{\partial P}{\partial T}\right)_\rho dT,
\end{equation}
we can also write a balance equation for pressure as,
\begin{equation}\label{eq:pressure_euler_CE}
    \partial_t^{(1)} P + \bm{u}\cdot\bm{\nabla}P + \rho c_s^2\bm{\nabla}\cdot\bm{u} = 0,
\end{equation}
where,
\begin{equation}
    c_s^2 = \left(\frac{\partial P}{\partial \rho}\right)_T + \frac{T}{c_v\rho^2} \left(\frac{\partial P}{\partial T}\right)_\rho^2.
\end{equation}

At order $\epsilon^2$, the continuity equation becomes
	\begin{equation}
	    \partial_t^{(2)}\rho = 0,
        \label{eq:NSFlevel_continuity1}
	\end{equation}
while for the momentum balance equation one has
    \begin{equation}
        \partial_t^{(2)}\left( \rho \bm{u} \right) + \bm{\nabla}\cdot \int\bm{v}\otimes\bm{v} f^{(1)} d\bm{v} = 0
        .
        \label{eq:approach2_momentum2_raw}
    \end{equation}
The second term can be expanded using the first-order-in-$\epsilon$ equation, as
\begin{equation}
    -\frac{1}{\tau} \int\bm{v}\otimes\bm{v} f^{(1)}d\bm{v} 
    + \int\bm{v}\otimes\bm{v} \mathcal{F}^{\star(1)}d\bm{v} 
    = \partial_t^{(1)} \int\bm{v}\otimes\bm{v}f^{(0)}d\bm{v} + \bm{\nabla}\cdot 
    \int\bm{v}\otimes\bm{v}\otimes\bm{v}f^{(0)}d\bm{v},
\end{equation}
where,
\begin{align}
    \partial_t^{(1)}\int\bm{v}\otimes\bm{v}f^{(0)}d\bm{v} &= \partial_t^{(1)}\left(\rho\bm{u}\otimes\bm{u} + P\bm{I}\right),\\
    \bm{\nabla}\cdot\int\bm{v}\otimes\bm{v}\otimes\bm{v}f^{(0)}d\bm{v} &= \bm{\nabla}\cdot\left(\rho\bm{u}\otimes\bm{u}\otimes\bm{u}\right) + \bm{\nabla}(P\bm{u})+\bm{\nabla}(P\bm{u})^\dagger + \bm{\nabla}\cdot(P\bm{u}),\\
    \int\bm{v}\otimes\bm{v}\mathcal{F}^{\star(1)}d\bm{v} &= \lambda\left(\bm{u}\otimes\bm{F}+\bm{F}\otimes\bm{u}\right) + \frac{ P}{\rho}\left(\frac{D+2}{D}-\frac{\rho c_s^2}{P}-\frac{\eta}{\mu}\right)\left(\bm{\nabla}\cdot\bm{u}\right),
\end{align}
The time derivative term can be expanded as,
\begin{equation}
    \partial_t^{(1)}\left(\rho\bm{u}\otimes\bm{u} + P\bm{I}\right) = 
    \bm{u}\otimes\partial_t^{(1)}\rho \bm{u} 
    + (\bm{u}\otimes\partial_t^{(1)}\rho \bm{u})^{\dagger} - \bm{u}\otimes\bm{u}\partial_t^{(1)}\rho + \partial_t^{(1)}P\bm{I}
    ,
\end{equation}
which then using Euler level balance equations, Eqs.~\eqref{eq:Eulerlevel_continuity1},~\eqref{eq:Eulerlevel_momentum1} and~\eqref{eq:pressure_euler_CE}, yields,
\begin{multline}
    \partial_t^{(1)}\left(\rho\bm{u}\otimes\bm{u} + P\bm{I}\right) = 
    -\bm{u}\otimes\left[\bm{\nabla}\cdot\rho\bm{u}\otimes\bm{u} + \bm{\nabla}P +\bm{F}\right]
    \\-\left(\bm{u}\otimes\left[\bm{\nabla}\cdot\rho\bm{u}\otimes\bm{u} + \bm{\nabla}P+ \bm{F}\right]\right)^{\dagger}
     + \bm{u}\otimes\bm{u} \bm{\nabla}\cdot\rho \bm{u} 
    - (\bm{\nabla} \cdot P \bm{u} )\bm{I} + \left(P-\rho c_s^2\right) (\bm{\nabla} \cdot \bm{u})\bm{I}
    .
\end{multline}
Adding contributions from other terms,
\begin{equation}
    \int \bm{v}\otimes\bm{v} f^{(1)}d\bm{v} = -\tau P\left[ \bm{\nabla}\bm{u} + \bm{\nabla}\bm{u}^{\dagger} +\left(\frac{\eta}{\mu}- \frac{2}{D}\right)(\bm{\nabla}\cdot\bm{u})\bm{I}\right]
    .
\end{equation}
Plugging this final expression into the momentum balance equation at order $\epsilon^2$, and setting $\tau = \mu/P$, results in
    \begin{equation}
        \partial_t^{(2)}(\rho \bm{u} )+ \bm{\nabla}\cdot \bm{T}_{\rm NS}  = 0
        .\label{eq:NSFlevel_momentum1}
    \end{equation}
For the energy balance at order $\epsilon^2$,
\begin{equation}
    \partial_t^{(2)} \rho E + \bm{\nabla}\cdot \int \bm{v} g^{(1)} d\bm{v} = 0,
\end{equation}
which can be evaluated using the order $\epsilon$ as,
\begin{equation}
    -\frac{1}{\tau} \int\bm{v} g^{(1)}d\bm{v} 
        + \int\bm{v}\mathcal{G}^{\star(1)} d\bm{v} 
        =
        \partial_t^{(1)}  \int\bm{v} g^{(0)}d\bm{v} 
        + \bm{\nabla}\cdot  \int\bm{v}\otimes\bm{v}g^{(0)}d\bm{v}  
        ,
\end{equation}
where,
\begin{align}
    \partial_t^{(1)}\int\bm{v}g^{(0)}d\bm{v} &= \partial_t^{(1)}\left(\rho E \bm{u} + P\bm{u}\right),\\
    \bm{\nabla}\cdot\int\bm{v}\otimes\bm{v}g^{(0)}d\bm{v} &= \bm{\nabla}\cdot\left(\bm{u}\otimes\bm{u}(\rho E+P)+(E+P/\rho)P\bm{I}\right),\\
    \int\bm{v}\mathcal{G}^{\star(1)}d\bm{v} &= \left(\bm{u}(\bm{u}\cdot\bm{F})+\bm{F}(E+P/\rho)\right) + P\bm{u}\left(\frac{D+2}{D}-\frac{\rho c_s^2}{P}-\frac{\eta}{\mu}\right)\left(\bm{\nabla}\cdot\bm{u}\right)\nonumber \\ &+P\left(\bm{\nabla}h - \frac{k}{\mu}\bm{\nabla}T\right).
\end{align}
Adding these terms up and using balance equations for $P\bm{u}$ and $\rho E \bm{u}$, 
\begin{equation}
    \partial_t^{(1)}P\bm{u} + \bm{\nabla}\cdot P\bm{u}\otimes\bm{u} + \frac{P}{\rho}\bm{\nabla}P +\frac{P}{\rho}\bm{\nabla}\cdot\bm{T}_K  + \bm{u}\left(\rho c_s^2 - P\right)\bm{\nabla}\cdot\bm{u} = 0,
\end{equation}
and
\begin{equation}
    \partial_t^{(1)}\rho E\bm{u} + \bm{\nabla}\cdot (\rho E + P)\bm{u}\otimes\bm{u} + E\bm{\nabla}P + E\bm{\nabla}\cdot\bm{T}_K  - P\bm{u}\cdot\bm{\nabla}\bm{u} + \bm{u}(\bm{u}\cdot\bm{\nabla}\cdot\bm{T}_K )= 0,
\end{equation}
we eventually recover,
\begin{equation}
    \partial_t^{(2)}\rho E + \bm{\nabla}\cdot \bm{u}\cdot\bm{T}_{\rm NS} - \bm{\nabla}\cdot k\bm{\nabla}T = 0.
\end{equation}
%%%%%%%%%%%%%%%%%%%%%%%%%%%%%%%%%%%%%%%%%%%%%%%%%%%%%%%%%%%%
\section{Second-order-in-time discretization\label{app:int_chars}}
We follow a procedure first introduced by \cite{he_1998_novel} to discretized the kinetic models introduced here,
\begin{align}
    \partial_t f&+\bm{v}\cdot\bm{\nabla}f=-\frac{1}{\tau}(f-f^{\rm eq})+\frac{1}{\lambda}(f_\lambda^\star-f^{\rm eq}),\\
    \partial_t g&+\bm{v}\cdot\bm{\nabla}g=-\frac{1}{\tau}(g-g^{\rm eq})+\frac{1}{\lambda}(g_\lambda^\star-g^{\rm eq}).
\end{align}
The main ingredient in space/time discretization is is the integration along characteristics, here the velocities $\bm{v}$, over a time $\delta t$ which leads to,
\begin{align}
    f(\bm{x}+\bm{v}\delta t, t+\delta t) &- f(\bm{x}, t) = \int_t^{t+\delta t}\left[\frac{1}{\tau}\left(f^{\rm eq} - f\right) + {\frac{1}{\lambda}}(f_\lambda^\star - f^{\rm eq})\right]dt',\\
    g(\bm{x}+\bm{v}\delta t, t+\delta t) &- g(\bm{x}, t) = \int_t^{t+\delta t}\left[\frac{1}{\tau}\left(g^{\rm eq} - g\right) + {\frac{1}{\lambda}}(g_\lambda^\star - g^{\rm eq})\right]dt'.
\end{align}
The integrals on the right hand sides are approximated using a trapezoidal rule,
\begin{multline}
    \int_t^{t+\delta t}\left[\frac{1}{\tau}\left(f^{\rm eq} - f\right) +{\frac{1}{\lambda}} (f_\lambda^\star - f^{\rm eq})\right]dt' = \frac{\delta t}{2\tau}\left(f^{\rm eq}(\bm{x},t) - f(\bm{x},t)\right) + \frac{\delta t}{2{\lambda}}\left(f_\lambda^\star(\bm{x},t) - f^{\rm eq}(\bm{x},t)\right) \\
    + \frac{\delta t}{2\tau}\left(f^{\rm eq}(\bm{x}+\bm{v}\delta t, t+\delta t) - f(\bm{x}+\bm{v}\delta t, t+\delta t)\right) \\ 
    + \frac{\delta t}{2{\lambda}}\left(f_\lambda^\star(\bm{x}+\bm{v}\delta t, t+\delta t) - f^{\rm eq}(\bm{x}+\bm{v}_i\delta t, t+\delta t)\right) + \mathcal{O}(\delta t^3).
 \end{multline}
A similar equation can be written for $g$ which we will omit for the sake of readability. The resulting system is implicit in time. To remove the implicitness, the following transformation of variables are introduced \citep{he_1998_novel},
\begin{align}
    \bar{f}(\bm{x},t) &= f(\bm{x},t) - \frac{\delta t}{2\tau}\left(f^{\rm eq}(\bm{x},t) - f(\bm{x},t)\right) - \frac{\delta t}{2{\lambda}}\left(f_\lambda^\star(\bm{x},t) - f^{\rm eq}(\bm{x},t)\right),\\
    \bar{g}(\bm{x},t) &= g(\bm{x},t) - \frac{\delta t}{2\tau}\left(g^{\rm eq}(\bm{x},t) - g(\bm{x},t)\right) - \frac{\delta t}{2{\lambda}}\left(g_\lambda^\star(\bm{x},t) - g^{\rm eq}(\bm{x},t)\right).
\end{align}
Introducing this transformation back into the integrated time-evolution equations,
\begin{align}
 \bar{f}(\bm{x}+\bm{v}\delta t, t+\delta t) &= \bar{f}(\bm{x}, t) + 2\beta\left(f^{\rm eq}(\bm{x},t) - \bar{f}(\bm{x},t)\right) + {\frac{\delta t}{\lambda}}\left(1-\beta\right) \left(f_\lambda^\star(\bm{x},t) - f^{\rm eq}(\bm{x},t)\right),\\
 \bar{g}(\bm{x}+\bm{v}\delta t, t+\delta t) &= \bar{g}(\bm{x}, t) + 2\beta\left(g^{\rm eq}(\bm{x},t) - \bar{g}(\bm{x},t)\right) + {\frac{\delta t}{\lambda}}\left(1-\beta\right) \left(g_\lambda^\star(\bm{x},t) - g^{\rm eq}(\bm{x},t)\right),
\end{align}
where $\beta\in[0,1]$ is the relaxation parameter,
\begin{equation}
    {\beta=\frac{\delta t}{2\tau+\delta t}.}
\end{equation}
%%%%%%%%%%%%%%%%%%%%%%%%%%
The final step is to evaluate moments of the distribution function that are needed to define $\{f^{\rm eq}, g^{\rm eq}\}$ and $\{f_\lambda^\star, g_\lambda^\star\}$ using the transformed distribution functions $\{\bar{f}, \bar{g}\}$. Integrating over $\{\bar{f}, \bar{g}\}$ and using the definitions for the transformed variables it is readily shown that,
\begin{align}
\int m \bar{f} d\bm{v} &= \int m f d\bm{v} = \rho,\\
\int m \bm{v} \bar{f} d\bm{v} &= \int m \bm{v} \left(f - \frac{\delta t}{2\lambda}\left(f_\lambda^\star-f^{\rm eq}\right)\right)d\bm{v} = \rho\bm{u}-\frac{\delta t}{2}\bm{F},\\
\int m \bar{g} d\bm{v} &= \int m\left(g - \frac{\delta t}{2\lambda}\left(g_\lambda^\star-g^{\rm eq}\right)\right)d\bm{v} = \rho E - \frac{\delta t}{2} \bm{u}\cdot\bm{F}.
\end{align}
This completes the integration along characteristics of the kinetic model.
{Renaming the variables $\{\bar{f},\bar{g}\}\to \{f,g\}$ and dropping the dependence on the untransformed distribution function, we obtain the second-order-in-time accurate kinetic equations \eqref{eq:pre_LBM_f} and \eqref{eq:pre_LBM_g} of the main text, along with the corresponding transform of the fields, Eqs. \eqref{eq:transform_rho}, \eqref{eq:transform_u} and \eqref{eq:transform_E}.}}
%}
\section{Multi-scale analysis of lattice Boltzmann model for compressible non-ideal flows\label{app:CE}}
The first step in the multi-scale analysis is a Taylor expansion of the lattice Boltzmann equations,
\begin{multline}
    \{f_i,g_i\}(\bm{x}+\bm{c}_i\delta t, t+\delta t)= \{f_i,g_i\}(\bm{x}, t) + 2\beta\left(\{f_i^{\rm eq},g_i^{\rm eq}\}(\bm{x},t) - \{f_i,g_i\}(\bm{x},t)\right) \\+ \left(1-\beta\right) \left(\{f_i^{\star},g_i^{\star}\}(\bm{x},t) - \{f_i^{\rm eq},g_i^{\rm eq}\}(\bm{x},t)\right),
\end{multline}
around $(\bm{x},t)$, leading to the following space and time-evolution equations,
\begin{multline}
    \delta t\mathcal{D}_t \{f_i,g_i\} + \frac{\delta t^2}{2}{\mathcal{D}_t}^2 \{f_i,g_i\} + \mathcal{O}\left(\delta t^3\right) = 2\beta\left(\{f_i^{\rm eq},g_i^{\rm eq}\} - \{f_i,g_i\}\right) \\ + \left(1-\beta\right)\left(\{f_i^\star,g_i^\star\} - \{f_i^{\rm eq},g_i^{\rm eq}\}\right).
\end{multline}
Introducing the flow characteristic size and time, $\mathcal{L}$ and $\mathcal{T}$ the equations are made non-dimensional as,
\begin{multline}
    \frac{\delta x}{\mathcal{L}}\mathcal{D}'_t \{f_i,g_i\} + \frac{\delta x^2}{2\mathcal{L}^2}{\mathcal{D}'_t}^2 \{f_i,g_i\} = 2\beta\left(\{f_i^{\rm eq},g_i^{\rm eq}\} - \{f_i,g_i\}\right) \\ + \left(1-\beta\right)\left(\{f_i^{\star},g_i^{\star}\} - \{f_i^{\rm eq},g_i^{\rm eq}\}\right),
\end{multline}
where,
\begin{equation}
    \mathcal{D}'_t = \frac{\mathcal{L}/\mathcal{T}}{\delta x/\delta t}\left(\partial'_t + \bm{c}'_i\cdot\bm{\nabla}'\right).
\end{equation}
Assuming acoustic scaling and hydrodynamic scaling, {$\varepsilon\sim\delta x/\mathcal{L} \sim\delta t/\mathcal{T}$} and dropping the primes,
\begin{equation}
    \varepsilon\mathcal{D}_t \{f_i,g_i\} + \frac{\varepsilon^2}{2}{\mathcal{D}_t}^2\{f_i,g_i\}  = 2\beta\left(\{f_i^{\rm eq},g_i^{\rm eq}\} - \{f_i,g_i\}\right)  + \left(1-\beta\right)\left(\{f_i^{\star},g_i^{\star}\} - \{f_i^{\rm eq},g_i^{\rm eq}\}\right),
\end{equation}
{
We introduce the following multi-scale expansions: 
\begin{subequations}
\begin{align}
    f_i &= f_i^{(0)} + \varepsilon f_i^{(1)} + \varepsilon^2 f_i^{(2)} + O(\varepsilon^3), \\
    g_i &= g_i^{(0)} + \varepsilon g_i^{(1)} + \varepsilon^2 g_i^{(2)} + O(\varepsilon^3), \\
    {f_i^\star} &= {f_i^\star}^{(0)} + \varepsilon {f_i^\star}^{(1)} + \varepsilon^2 {f_i^\star}^{(2)} + O(\varepsilon^3), \\
    {g_i^\star} &= {g_i^\star}^{(0)} + \varepsilon {g_i^\star}^{(1)} + \varepsilon^2 {g_i^\star}^{(2)} + O(\varepsilon^3), \\
    \partial_t &= \varepsilon \partial_t^{(1)} + \varepsilon^2 \partial_t^{(2)} + O(\varepsilon^3).
\end{align}
\end{subequations}
Noting that for the definition of ${f_i^\star}$ in \eqref{eq:LBMstar} divergence from the equilibrium will arise only at the $\varepsilon^1$ level (i.e. ${f_i^\star}^{(0)} = f^{\rm eq}_i$), we separate by orders of the smallness parameter,
}
	\begin{subequations}
	\begin{align}
    \varepsilon^0 &: \{f_i^{(0)},g_i^{(0)}\} = \{f_i^{\rm eq},g_i^{\rm eq}\},\\
	\varepsilon &: \mathcal{D}_{t}^{(1)} \{f_i^{(0)},g_i^{(0)}\} = -2\beta \{f_i^{(1)},g_i^{(1)}\} + \left(1-\beta\right)\{{f^\star}_i^{(1)},{g^\star}_i^{(1)}\},\label{eq:CE_order_1}\\
	\varepsilon^2 &: \partial_t^{(2)}\{f_i^{(0)},g_i^{(0)}\} + \mathcal{D}_{t}^{(1)}(1-\beta) \left(\{f_i^{(1)},g_i^{(1)}\} + {\frac{1}{2}} \{{f^\star}_i^{(1)},{g^\star}_i^{(1)}\}\right) =  \nonumber \\ &-2\beta \{f_i^{(2)},g_i^{(2)}\} + \left(1-\beta\right)\{{f^\star}_i^{(2)},{g^\star}_i^{(2)}\}.
	\end{align}
    \label{Eq:CE_Eq_orders_LB}
    \end{subequations}
The following solvability conditions apply,
    \begin{eqnarray}
		\sum_{i=1}^Q {f}_i^{(k)} = 0,\ \forall k>0, \label{Eq:CE_solvability_LB_1}\\
		\sum_{i=1}^Q \bm{c}_i {f}_i^{(1)}  + \frac{1}{2}\sum_{i=1}^Q \bm{c}_i {f^\star}_i^{(1)}  = 0, \label{Eq:CE_solvability_LB_2}\\
        \sum_{i=1}^Q \bm{c}_i {f}_i^{(k)}  = 0,\ \forall k>1, \label{Eq:CE_solvability_LB_3}\\
        \sum_{i=1}^Q {g}_i^{(1)} + \frac{1}{2} \sum_{i=1}^Q {g^\star}_i^{(1)} = 0,
	    \label{Eq:CE_solvability_LB_4}\\
        \sum_{i=1}^Q {g}_i^{(k)} = 0,\ \forall k>0.
	    \label{Eq:CE_solvability_LB_5}
    \end{eqnarray}
Taking the zeroth-order moment of Eq.~\eqref{eq:CE_order_1} {for $f_i$},
\begin{equation}
    \partial_t^{(1)}\rho + \bm{\nabla}\cdot\rho\bm{u} = 0,
\end{equation}
where we have used,
\begin{equation}
    \sum_{i=1}^{Q} {f_i^\star}^{(1)} = 0,
\end{equation}
and solvability condition \eqref{Eq:CE_solvability_LB_1}. For the first-order moment of \eqref{eq:CE_order_1} {for $f_i$},
\begin{equation}
    \partial_t^{(1)}\rho\bm{u} + \bm{\nabla}\cdot\rho\bm{u}\otimes\bm{u} + \bm{\nabla} P = -2\beta\overbrace{\left(\sum_{i=1}^Q \bm{c}_if_i^{(1)}+\frac{1}{2}\bm{c}_i{f_i^\star}^{(1)}\right)}^{=0}  + \underbrace{\sum_{i=1}^Q \bm{c}_i{f_i^\star}^{(1)}}_{\bm{F}},
\end{equation}
where we used \eqref{Eq:CE_solvability_LB_2} and,
\begin{equation}
    \bm{F} = -\bm{\nabla}\cdot T_K,
\end{equation}
where,
\begin{equation}
    T_K = \kappa \bm{\nabla}\rho\otimes\bm{\nabla}\rho - \kappa\left(\rho \bm{\nabla}^2\rho+\frac{1}{2}{\lvert \bm{\nabla}\rho\lvert}^2\right)\bm{I}.
\end{equation}
The force can also be shown to simplify to,
\begin{equation}
    \bm{F} = \kappa \rho \bm{\nabla} \bm{\nabla}^2\rho.
\end{equation}
Finally taking the zeroth-order moment for $g_i$,
\begin{equation}
    \partial_t^{(1)}\rho E + \bm{\nabla}\cdot \bm{u}\left(\rho E + P\right) = -2\beta\overbrace{\left(\sum_{i=1}^Q g_i^{(1)}+\frac{1}{2}{g_i^\star}^{(1)}\right)}^{=0} + \underbrace{\sum_{i=1}^Q {g_i^\star}^{(1)}}_{\bm{u}\cdot\bm{F}}.
\end{equation}
Summing up balance equations at order $\varepsilon$,
	\begin{eqnarray}
	    \partial_t^{(1)}\rho + \bm{\nabla}\cdot \rho \bm{u} &=& 0,\label{eq:LB_Euler_density_balance}\\
	    \partial_t^{(1)}\rho \bm{u} + \bm{\nabla}\cdot \left(\rho \bm{u}\otimes\bm{u} + P\bm{I} \right) - \bm{F} &=& 0,\label{eq:LB_Euler_momentum_balance}\\
        \partial_t^{(1)}\rho E + \bm{\nabla}\cdot \rho \bm{u}\left(E + P/\rho\right) - \bm{u}\cdot\bm{F} &=& 0. \label{eq:LB_Euler_energy_balance}
	\end{eqnarray}
The last equation Eq.~\eqref{eq:LB_Euler_energy_balance} can be transformed into a balance equation for internal energy, using ,
\begin{equation}
    \partial_t^{(1)} \mathcal{K} + \bm{\nabla}\cdot(\bm{u} \mathcal{K}) + \bm{u}\cdot\bm{\nabla}P - \bm{u}\cdot\bm{F} = 0,
\end{equation}
as,
{\begin{equation}
    \partial_t^{(1)} \rho e + \bm{\nabla}\cdot\rho \bm{u} e + P\bm{\nabla}\cdot\bm{u}  = 0.
\end{equation}}
Furthermore, using
\begin{equation}
    de = c_v dT - \left(T \left(\frac{\partial P}{\partial T}\right)_\rho- P\right)\frac{d\rho}{\rho^2},
\end{equation}
and Eq.~\eqref{eq:LB_Euler_density_balance} a balance equation for temperature can be derived as,
\begin{equation}
    \partial_t^{(1)}T + \bm{u}\cdot\bm{\nabla}T + \frac{T}{\rho c_v}\left(\frac{\partial P}{\partial T}\right)_\rho\bm{\nabla}\cdot\bm{u} = 0.
\end{equation}
Finally, using
\begin{equation}
    dP = \left(\frac{\partial P}{\partial \rho}\right)_T d\rho + \left(\frac{\partial P}{\partial T}\right)_\rho dT,
\end{equation}
we can also write a balance equation for pressure as,
\begin{equation}
    \partial_t^{(1)} P + \bm{u}\cdot\bm{\nabla}P + \rho c_s^2\bm{\nabla}\cdot\bm{u} = 0,
\end{equation}
where,
\begin{equation}
    c_s^2 = \left(\frac{\partial P}{\partial \rho}\right)_T + \frac{T}{c_v\rho^2} \left(\frac{\partial P}{\partial T}\right)_\rho^2.
\end{equation}

At order $\varepsilon^2$, the zeroth order moment {of $f_i$} leads to,
\begin{equation}
\partial_t^{(2)} \rho = 0,
\end{equation}
{and} the first order moments,
\begin{equation}
    \partial_t^{(2)}\rho\bm{u} + \bm{\nabla}\cdot\left(1-\beta\right)\left[\left(\sum_{i=1}^Q \bm{c}_i\otimes\bm{c}_i {f_i}^{(1)}\right)  + \frac{1}{2}\left(\sum_{i=1}^Q \bm{c}_i\otimes\bm{c}_i {f^\star}_i^{(1)}\right)\right] = 0.
\end{equation}
Here we can use Eq.~\eqref{eq:CE_order_1} to obtain,
\begin{equation}
    \partial_t^{(2)}\rho\bm{u} + \bm{\nabla}\cdot\left(\frac{1}{2}-\frac{1}{2\beta}\right)\left(\partial_t^{(1)}\Pi_2(f^{(0)}) + \bm{\nabla}\Pi_3(f^{(0)})  - \sum_{i=1}^Q \bm{c}_i\otimes\bm{c}_i {f^\star}_i^{(1)}\right) = 0.\label{eq:epsilon2_mom_1_1}
\end{equation}
Here, $\Pi^{(0)}_2(f)$ and $\Pi^{(0)}_3(f)$ are the second and third order moment tensor of the equilibrium distribution function,
	\begin{eqnarray}
	    \Pi_2(f^{(0)}) &=& \rho\bm{u}\otimes\bm{u} + P\bm{I},\\
	    \Pi_3(f^{(0)}) &=& \rho\bm{u}\otimes(\bm{u}\otimes\bm{u}+P\bm{I})\circ(\bm{1}-\bm{J}) + 3\rho \varsigma^2 \bm{u}\otimes\bm{I}\circ\bm{J}. 
	\end{eqnarray}
Using Eqs.~\eqref{eq:LB_Euler_density_balance} and \eqref{eq:LB_Euler_momentum_balance} we can write,
\begin{multline}
    \partial^{(1)}_t \left(\rho\bm{u}\otimes\bm{u} + P\bm{I}\right) = \bm{u}\otimes\bm{F} + \bm{F}\otimes\bm{u}  -\bm{\nabla}\cdot\rho\bm{u}\otimes\bm{u}\otimes\bm{u} - \bm{\nabla}P\bm{u} - (\bm{\nabla}P\bm{u})^\dagger \\  P\left(\bm{\nabla}\bm{u} + \bm{\nabla}\bm{u}^\dagger\right)  + \partial^{(1)}_t P \bm{I}.
\end{multline}
For the last term, i.e. $\partial^{(1)}_t P$, we write a balance equation for $P$,
\begin{equation}
    \partial_t^{(1)} P  = \left(P - \rho c_s^2\right)\bm{\nabla}\cdot\bm{u} - \bm{\nabla}\cdot P\bm{u},
\end{equation}
while,
\begin{equation}
    \bm{\nabla}\cdot\Pi_3(f^{(0)}) = \left[\bm{\nabla}\cdot\rho\bm{u}\otimes\bm{u}\otimes\bm{u} + \bm{\nabla}P\bm{u} + \bm{\nabla}P\bm{u}^\dagger\right] + \rho\Psi +\bm{I}\bm{\nabla}\cdot P\bm{u},
\end{equation}
where,
\begin{equation}
    \Psi_{\alpha\alpha} = -\frac{1}{\rho}\partial_\alpha u_\alpha\left(u_\alpha^2 + 3(P-\rho\varsigma^2)\right).
\end{equation}
Adding up all the terms,
\begin{multline}
    \partial_t^{(1)}\Pi_2(f^{(0)}) + \bm{\nabla}\cdot\Pi_3(f^{(0)}) = \bm{u}\otimes\bm{F} + {\bm{F}\otimes\bm{u}} + P\left(\bm{\nabla}\bm{u} + \bm{\nabla}\bm{u}^\dagger\right) \\+ \left(P - \rho c_s^2\right)\bm{\nabla}\cdot\bm{u}\bm{I} - \rho\Psi.
\end{multline}
Setting,
\begin{equation}
    \sum_{i=1}^Q \bm{c}_i\otimes\bm{c}_i {f^\star}_i^{(1)} = \left(\bm{u}\otimes\bm{F} + \bm{F}\otimes\bm{u} + \rho\Psi + P\left(\frac{D+2}{D}-\frac{\rho c_s^2}{P} -\frac{\eta}{\mu}\right)\bm{\nabla}\cdot\bm{u}\bm{I}\right) 
\end{equation}
and plugging it back into Eq.~\eqref{eq:epsilon2_mom_1_1},
\begin{equation}
    \partial_t^{(2)}\rho\bm{u} - \bm{\nabla}\cdot\left[ \mu\left(\bm{\nabla}\bm{u} + \bm{\nabla}\bm{u}^\dagger - \frac{2}{D}\bm{\nabla}\cdot\bm{u} \bm{I}\right) + \eta \bm{\nabla}\cdot\bm{u}\bm{I}\right]  = 0,
\end{equation}
Where we used,
\begin{equation}
    \mu = \left(\frac{1}{2\beta}-\frac{1}{2}\right) P \delta t.
\end{equation}
For the second population, at order $\varepsilon^2$,
\begin{equation}
    \partial_t^{(2)}\rho E + \bm{\nabla}\cdot\left(\frac{1}{2}-\frac{1}{2\beta}\right)\left( \partial_t^{(1)}\Pi_1(g_i^{(0)}) + \bm{\nabla}\cdot\Pi_2(g_i^{(0)})  - \sum_{i=1}^Q \bm{c}_i {g_i^\star}^{(1)}\right) = 0,
\end{equation}
where,
\begin{eqnarray}
    \Pi_1(g_i^{(0)}) &=& \bm{u}(\rho E + P),\\
    \Pi_2(g_i^{(0)}) &=& \bm{u}\otimes\bm{u}\left(\rho E+2P\right) + \left(E + P/\rho\right)P\bm{I}.
\end{eqnarray}
Here we can use,
\begin{equation}\label{eq:pu_balance}
    \partial_t^{(1)}P\bm{u} + \bm{\nabla}\cdot P\bm{u}\otimes\bm{u} + \frac{P}{\rho}\bm{\nabla}P -\frac{P}{\rho}\bm{F} + \bm{u}\left(\rho c_s^2 - P\right)\bm{\nabla}\cdot\bm{u} = 0,
\end{equation}
and
\begin{equation}\label{eq:Eu_balance}
    \partial_t^{(1)}\rho E\bm{u} + \bm{\nabla}\cdot (\rho E + P)\bm{u}\otimes\bm{u} + E\bm{\nabla}P - E\bm{F} - P\bm{u}\cdot\bm{\nabla}\bm{u} - \bm{u}(\bm{u}\cdot\bm{F})= 0,
\end{equation}
Adding up both contributions,
\begin{multline}
    \partial_t^{(1)}\Pi_1(g_i^{(0)}) + \bm{\nabla}\cdot\Pi_2(g_i^{(0)}) = \underbrace{-2\bm{\nabla}\cdot P \bm{u}\otimes\bm{u} + 2\bm{\nabla}\cdot P \bm{u}\otimes\bm{u}}_{=0} \underbrace{+ \bm{\nabla}\cdot \rho E \bm{u}\otimes\bm{u} - \bm{\nabla}\cdot \rho E \bm{u}\otimes\bm{u}}_{=0}
     \\ \underbrace{+ \bm{\nabla} P(P/\rho + E) - (E+P/\rho)\bm{\nabla}P}_{P\bm{\nabla}(E+P/\rho)} + P\bm{u}\cdot\bm{\nabla}\bm{u} - \bm{u}\left(\rho c_s^2 - P\right)\bm{\nabla}\cdot\bm{u}
     \\+(P/\rho+E)\bm{F} + \bm{u}(\bm{u}\cdot\bm{F}).
\end{multline}
Further expanding,
\begin{multline}
    \partial_t^{(1)}\Pi_1(g_i^{(0)}) + \bm{\nabla}\cdot\Pi_2(g_i^{(0)}) = 
     P\bm{\nabla}h \overbrace{+ P\bm{\nabla}(\bm{u}^2/2) + P\bm{u}\cdot\bm{\nabla}\bm{u}}^{=P\bm{u}\cdot(\bm{\nabla}\bm{u} + \bm{\nabla}\bm{u}^\dagger)} + \bm{u}\left(P - \rho c_s^2\right)\bm{\nabla}\cdot\bm{u}
     \\ +(P/\rho+E)\bm{F} + \bm{u}(\bm{u}\cdot\bm{F}),
\end{multline}
where $h = e + P$. Plugging this back into the balance equation,
\begin{multline}
    \partial_t^{(2)}\rho E - \bm{\nabla}\cdot\underbrace{\bm{u}\cdot\left[\mu\left(\bm{\nabla}\bm{u} + \bm{\nabla}\bm{u} - \frac{2}{D}\bm{\nabla}\cdot\bm{u}\bm{I}\right) + \eta\bm{\nabla}\cdot\bm{u}\bm{I}\right]}_{-\bm{u}\cdot\bm{T}_{\rm NS}} \\ + \bm{\nabla}\cdot\left(\frac{1}{2}-\frac{1}{2\beta}\right)\left(P\bm{\nabla}h  + \bm{u}\cdot P\left(\frac{D+2}{D}-\frac{\rho c_s^2}{P}-\frac{\eta }{\mu}\right)\bm{\nabla}\cdot\bm{u}\bm{I}
    \right. \\ \left. + (P/\rho+E)\bm{F} + \bm{u}(\bm{u}\cdot\bm{F}) 
    - \sum_{i=1}^Q \bm{c}_i {g_i^{\star}}^{(1)}\right) = 0.
\end{multline}
Using the definition of the shifted equilibrium,
\begin{eqnarray}
    \Pi_0(g_i^{\star(1)}) &=& \bm{u}\cdot\bm{F},\\
    \Pi_1(g_i^{\star(1)}) &=& \left(\bm{u}(\bm{u}\cdot\bm{F}) + \bm{F} \left(\frac{P}{\rho} + E\right) \right. \nonumber \\ & & \left. + P\left(\bm{\nabla}h - \frac{k}{\mu}\bm{\nabla}T + \bm{u} \cdot\left(\frac{D+2}{D}-\frac{\rho c_s^2}{P}-\frac{\eta}{\mu}\right)\bm{\nabla}\cdot\bm{u}\bm{I}\right)\right).
\end{eqnarray}
we recover,
\begin{equation}
    \partial_t^{(2)}\rho E + \bm{\nabla}\cdot \bm{u}\cdot\bm{T}_{\rm NS} - \bm{\nabla}\cdot k\bm{\nabla}T = 0.
\end{equation}
\section{Convergence study of shock-liquid column interaction}
\begin{figure}
	\centering
	\includegraphics[width=0.6\linewidth,keepaspectratio]{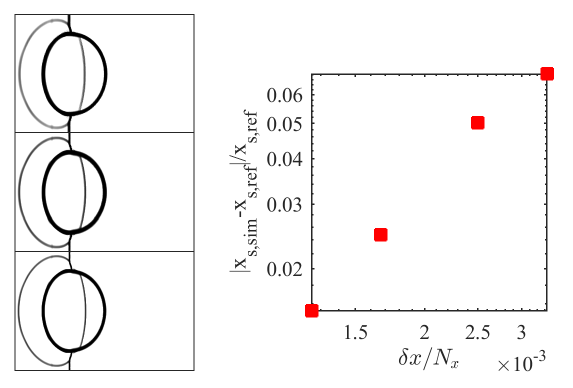}
	\caption{{Left: Numerical Schlieren image of shock-liquid column interaction at ${\rm Ma}=1.47$ with resolutions (from top to bottom): $600\times600$, $800\times800$, $1000\times1000$. Right: error in position of shock inside the liquid column along the center-line $x_s$.}}
	\label{Fig:shock_drop_converge_147}
\end{figure}

\begin{figure}
	\centering
	\includegraphics[width=0.6\linewidth,keepaspectratio]{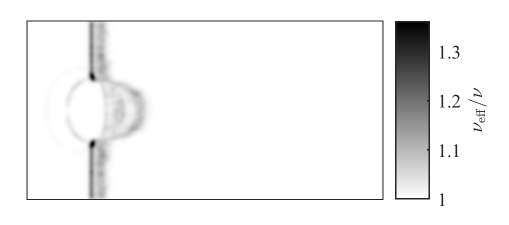}
	\caption{{Distribution of normalized effective viscosity in the domain at $t/t_0=0.3$ for ${\rm Ma}=1.47$ and resolution of $800\times800$.}}
	\label{Fig:shock_drop_num_diss_147}
\end{figure}

{The simulations reported for the shock-column interaction were run with a resolution of $800\times800$. To check the convergence behaviour for this configuration simulations were run with resolutions from $300\times300$ up to $1000\times1000$ using acoustic scaling for the time-step. The position of the shock inside the liquid column at $t/t_0=0.3$ was then extracted for each run and the relative error with respect to the highest resolution simulation measured. The scaling of this error with resolution along with a visual illustration of the field for ${\rm Ma}=1.47$ are shown in Fig.~\ref{Fig:shock_drop_converge_147}.}
{Fitting the four data points obtained with simulations a slope $1.56$ was obtained. Note that in all simulations the shock-capturing non-linear numerical dissipation of Eqs.~\eqref{eq:shock_cap} and \eqref{eq:shock_cap_k} was on, showing that it does not diminish the overall accuracy of the solver. To better illustrate the operation mode of the numerical dissipation, the normalized effective viscosity in the domain at $t/t_0=0.3$ for ${\rm Ma}=1.47$ and resolution of $800\times800$ is shown in Fig.~\ref{Fig:shock_drop_num_diss_147}.}
\bibliographystyle{jfm}
\bibliography{References}
\end{document}